\documentclass[12pt]{article}
\usepackage[dvips]{graphicx}
\usepackage{float}
\usepackage{epsfig}
\usepackage{ulem}
\usepackage{latexsym,amsmath,amsfonts,amssymb}
\usepackage[latin1]{inputenc}
\usepackage{rotating}
\usepackage[american]{babel}
\usepackage[dvips]{graphicx}
\usepackage{bbm}
\usepackage{color}
\usepackage{slashed}
\usepackage[unicode]{hyperref}
\usepackage{lscape}
\usepackage{bigints}
\usepackage{enumerate}
\usepackage[shortlabels]{enumitem}
\usepackage{tikz}
\usetikzlibrary{decorations.pathreplacing}
\usetikzlibrary{shapes}
\def\im{Invent. Math.}

\def\hat{\widehat}
\def\a{\alpha}
\def\b{\beta}
\def\c{\gamma}
\def\d{\delta}
\def\eps{\epsilon}           
\def\f{\phi}               
\def\vf{\varphi}  
\def\tvf{\tilde{\varphi}}
\def\vp{\varphi}
\def\g{\gamma}
\def\h{\eta}
\def\j{\psi}
\def\k{\kappa}                    
\def\l{\lambda}
\def\m{\mu}
\def\n{\nu}
\def\o{\omega}  \def\w{\omega}

\def\q{\theta}  \def\th{\theta}                  
\def\r{\rho}                                     
\def\s{\sigma}                                   
\def\t{\tau}
\def\u{\upsilon}
\def\x{\xi}
\def\z{\zeta}
\def\pt{\tilde{\varphi}}
\def\lab{\label}
\def\6{\partial}
\def\wg{\wedge}
\def\bpsi{\bar{\psi}}
\def\bt{\bar{\theta}}
\def\bvf{\bar{\varphi}}

\DeclareMathOperator{\tr}{tr}

\newcommand{\be}{\begin{equation}}
\newcommand{\ee}{\end{equation}}
\newcommand{\beq}{\begin{equation}}
\newcommand{\eeq}{\end{equation}}
\newcommand{\bea}{\begin{eqnarray}}
\newcommand{\eea}{\end{eqnarray}}

\newcommand{\ba}{\begin{eqnarray}}
\newcommand{\ea}{\end{eqnarray}}

\newcommand{\beqs}{\begin{eqnarray}}
\newcommand{\eeqs}{\end{eqnarray}}
\newcommand{\bal}{\begin{aligned}}
\newcommand{\eal}{\end{aligned}}
\makeatletter
\newcommand\setItemnumber[1]{\setcounter{enum\romannumeral\@enumdepth}{\numexpr#1-1\relax}}
\makeatother
\begin{document}
\baselineskip=15.5pt
\pagestyle{plain}
\setcounter{page}{1}

\def\del{{\partial}}
\def\vev#1{\left\langle #1 \right\rangle}
\def\cn{{\cal N}}
\def\co{{\cal O}}


\def\IC{{\mathbb C}}
\def\IR{{\mathbb R}}
\def\IZ{{\mathbb Z}}
\def\RP{{\bf RP}}
\def\CP{{\bf CP}}
\def\Poincaré{{Poincar\'e }}
\def\tr{{\rm tr}}
\def\tp{{\tilde \Phi}}

\def\TL{\hfil$\displaystyle{##}$}
\def\TR{$\displaystyle{{}##}$\hfil}
\def\TC{\hfil$\displaystyle{##}$\hfil}
\def\TT{\hbox{##}}
\def\HLINE{\noalign{\vskip1\jot}\hline\noalign{\vskip1\jot}}
\def\seqalign#1#2{\vcenter{\openup1\jot
   \halign{\strut #1\cr #2 \cr}}}
\def\lbldef#1#2{\expandafter\gdef\csname #1\endcsname {#2}}
\def\eqn#1#2{\lbldef{#1}{(\ref{#1})}%
\begin{equation} #2 \label{#1} \end{equation}}
\def\eqalign#1{\vcenter{\openup1\jot
     \halign{\strut\span\TL & \span\TR\cr #1 \cr
    }}}

\def\eno#1{(\ref{#1})}
\def\half{\frac{1}{2}}



\def\ads{{\it AdS}}
\def\adsp{{\it AdS}$_{p+2}$}
\def\cft{{\it CFT}}

\newcommand{\ber}{\begin{eqnarray}}
\newcommand{\eer}{\end{eqnarray}}

\newcommand{\beqar}{\begin{eqnarray}}
\newcommand{\cN}{{\cal N}}
\newcommand{\cO}{{\cal O}}
\newcommand{\cA}{{\cal A}}
\newcommand{\cT}{{\cal T}}
\newcommand{\cF}{{\cal F}}
\newcommand{\cC}{{\cal C}}
\newcommand{\cR}{{\cal R}}
\newcommand{\cW}{{\cal W}}
\newcommand{\eeqar}{\end{eqnarray}}
\newcommand{\tht}{\thteta}
\newcommand{\lm}{\lambda}\newcommand{\Lm}{\Lambda}


\newcommand{\nonu}{\nonumber}
\newcommand{\oh}{\displaystyle{\frac{1}{2}}}
\newcommand{\dsl}
   {\kern.06em\hbox{\raise.15ex\hbox{$/$}\kern-.56em\hbox{$\partial$}}}
\newcommand{\id}{i\!\!\not\!\partial}
\newcommand{\as}{\not\!\! A}
\newcommand{\ps}{\not\! p}
\newcommand{\ks}{\not\! k}
\newcommand{\D}{{\cal{D}}}
\newcommand{\dv}{d^2x}
\newcommand{\Z}{{\cal Z}}
\newcommand{\N}{{\cal N}}
\newcommand{\Dsl}{\not\!\! D}
\newcommand{\Bsl}{\not\!\! B}
\newcommand{\Psl}{\not\!\! P}

\newcommand{\eeqarr}{\end{eqnarray}}
\newcommand{\ZZ}{{\rm \kern 0.275em Z \kern -0.92em Z}\;}


\def\del{{\delta^{\hbox{\sevenrm B}}}} \def\ex{{\hbox{\rm e}}}
\def\azb{A_{\bar z}} \def\az{A_z} \def\bzb{B_{\bar z}} \def\bz{B_z}
\def\czb{C_{\bar z}} \def\cz{C_z} \def\dzb{D_{\bar z}} \def\dz{D_z}
\def\im{{\hbox{\rm Im}}} \def\mod{{\hbox{\rm mod}}} \def\tr{{\hbox{\rm Tr}}}
\def\ch{{\hbox{\rm ch}}} \def\imp{{\hbox{\sevenrm Im}}}
\def\trp{{\hbox{\sevenrm Tr}}} \def\vol{{\hbox{\rm Vol}}}
\def\rl{\Lambda_{\hbox{\sevenrm R}}} \def\wl{\Lambda_{\hbox{\sevenrm W}}}
\def\fc{{\cal F}_{k+\cox}} \def\vev{vacuum expectation value}
\def\nodiv{\mid{\hbox{\hskip-7.8pt/}}}
\def\ie{{\em i.e.}}
\def\ie{\hbox{\it i.e.}}

\def\CC{{\mathchoice
{\rm C\mkern-8mu\vrule height1.45ex depth-.05ex
width.05em\mkern9mu\kern-.05em}
{\rm C\mkern-8mu\vrule height1.45ex depth-.05ex
width.05em\mkern9mu\kern-.05em}
{\rm C\mkern-8mu\vrule height1ex depth-.07ex
width.035em\mkern9mu\kern-.035em}
{\rm C\mkern-8mu\vrule height.65ex depth-.1ex
width.025em\mkern8mu\kern-.025em}}}

\def\RR{{\rm I\kern-1.6pt {\rm R}}}
\def\NN{{\rm I\!N}}
\def\ZZ{{\rm Z}\kern-3.8pt {\rm Z} \kern2pt}
\def\IB{\relax{\rm I\kern-.18em B}}
\def\ID{\relax{\rm I\kern-.18em D}}
\def\II{\relax{\rm I\kern-.18em I}}
\def\IP{\relax{\rm I\kern-.18em P}}
\newcommand{\CS}{{\scriptstyle {\rm CS}}}
\newcommand{\CSs}{{\scriptscriptstyle {\rm CS}}}
\newcommand{\rc}{\nonumber\\}
\newcommand{\bear}{\begin{eqnarray}}
\newcommand{\eear}{\end{eqnarray}}

\newcommand{\LL}{{\cal L}}

\def\mani{{\cal M}}
\def\calo{{\cal O}}
\def\calb{{\cal B}}
\def\calw{{\cal W}}
\def\calz{{\cal Z}}
\def\cald{{\cal D}}
\def\calc{{\cal C}}

\def\to{\rightarrow}
\def\ele{{\hbox{\sevenrm L}}}
\def\ere{{\hbox{\sevenrm R}}}
\def\zb{{\bar z}}
\def\wb{{\bar w}}
\def\nodiv{\mid{\hbox{\hskip-7.8pt/}}}
\def\menos{\hbox{\hskip-2.9pt}}
\def\dr{\dot R_}
\def\drr{\dot r_}
\def\ds{\dot s_}
\def\da{\dot A_}
\def\dga{\dot \gamma_}
\def\ga{\gamma_}
\def\dal{\dot\alpha_}
\def\al{\alpha_}
\def\cl{{closed}}
\def\cls{{closing}}
\def\vev{vacuum expectation value}
\def\tr{{\rm Tr}}
\def\to{\rightarrow}
\def\too{\longrightarrow}


\def\a{\alpha}
\def\b{\beta}
\def\c{\gamma}
\def\d{\delta}
\def\e{\epsilon}           
\def\F{\Phi}
\def\f{\phi}               
\def\vf{\varphi}  \def\tvf{\tilde{\varphi}}
\def\vp{\varphi}
\def\g{\gamma}
\def\h{\eta}
\def\j{\psi}
\def\k{\kappa}                    
\def\l{\lambda}
\def\m{\mu}
\def\n{\nu}
\def\o{\omega}  \def\w{\omega}
\def\q{\theta}  \def\th{\theta}                  
\def\r{\rho}                                     
\def\s{\sigma}                                   
\def\t{\tau}
\def\u{\upsilon}
\def\x{\xi}
\def\X{\Xi}
\def\z{\zeta}
\def\pt{\tilde{\varphi}}
\def\lab{\label}
\def\6{\partial}
\def\wg{\wedge}
\def\atanh{{\rm arctanh}}
\def\bpsi{\bar{\psi}}
\def\bt{\bar{\theta}}
\def\bvf{\bar{\varphi}}

%



\newfont{\namefont}{cmr10}
\newfont{\addfont}{cmti7 scaled 1440}
\newfont{\boldmathfont}{cmbx10}
\newfont{\headfontb}{cmbx10 scaled 1728}





\newcommand{\re}{\,\mathbb{R}\mbox{e}\,}
\newcommand{\hyph}[1]{$#1$\nobreakdash-\hspace{0pt}}
\providecommand{\abs}[1]{\lvert#1\rvert}
\newcommand{\Nugual}[1]{$\mathcal{N}= #1 $}
\newcommand{\sub}[2]{#1_\text{#2}}
\newcommand{\partfrac}[2]{\frac{\partial #1}{\partial #2}}
\newcommand{\bsp}[1]{\begin{equation} \begin{split} #1 \end{split} \end{equation}}
\newcommand{\calF}{\mathcal{F}}
\newcommand{\calO}{\mathcal{O}}
\newcommand{\calM}{\mathcal{M}}
\newcommand{\calV}{\mathcal{V}}
\newcommand{\bbZ}{\mathbb{Z}}
\newcommand{\bbC}{\mathbb{C}}
\newcommand{\cK}{{\cal K}}
\newcommand{\gt}{\tilde{g}}

\newcommand{\Thq}{\Theta\left(\r-\r_q\right)}
\newcommand{\Dq}{\d\left(\r-\r_q\right)}
\newcommand{\kten}{\kappa^2_{\left(10\right)}}
\newcommand{\pbi}[1]{\imath^*\left(#1\right)}
\newcommand{\ho}{\hat{\omega}}
\newcommand{\tth}{\tilde{\th}}
\newcommand{\tf}{\tilde{\f}}
\newcommand{\tj}{\tilde{\j}}
\newcommand{\tw}{\tilde{\omega}}
\newcommand{\tz}{\tilde{z}}
\newcommand{\prj}[2]{(\partial_r{#1})(\partial_{\j}{#2})-(\partial_r{#2})(\partial_{\j}{#1})}
\def\atanh{{\rm arctanh}}
\def\sech{{\rm sech}}
\def\csch{{\rm csch}}
\allowdisplaybreaks[1]

\newcommand \arXiv [1]{\href{http://arxiv.org/abs/#1}{\tt arXiv:#1}} 

\def\red{\textcolor[rgb]{0.98,0.00,0.00}}

\newcommand{\Dan}[1] {{\textcolor{blue}{#1}}}

\numberwithin{equation}{section}

\newcommand{\Tr}{\mbox{Tr}}    


%

\setcounter{footnote}{0}
\renewcommand{\theequation}{{\rm\thesection.\arabic{equation}}}

\begin{titlepage}

\begin{center}

\vskip .5in 
\noindent

{\Large \bf{ Holographic description of SCFT$_5$ compactifications} }
\bigskip\medskip
\\
 Andrea Legramandi\footnote{andrea.legramandi@swansea.ac.uk} and Carlos Nunez\footnote{c.nunez@swansea.ac.uk}
\\

\bigskip\medskip
{\small 

Department of Physics, Swansea University, Swansea SA2 8PP, United Kingdom}

\vskip .5cm 
\vskip .9cm 
     	\abstract{
	We present three infinite families of supersymmetric Type IIB backgrounds with AdS$_4$, AdS$_3$ and AdS$_2$ factors, dual to SCFTs in $3$, $2$ and $1$ space-time dimensions respectively. These field theories emerge at low energies, after a twisted compactification  of a family of five dimensional ${\cal N}=2$ SCFTs on hyperbolic spaces. The holographic flows across dimensions are explicitly computed. We also discuss a family of SUSY breaking backgrounds, dual to a QCD-like quiver with massive (bi)fundamental matter. Some field theoretical observables are computed for these  theories at the fixed points and along the flow.}
	
	\end{center}

\noindent

\noindent
\vskip .5cm
\vskip .5cm
\vfill
\eject

\end{titlepage}

\setcounter{footnote}{0}


\normalsize

%
\renewcommand{\theequation}{{\rm\thesection.\arabic{equation}}}

\tableofcontents

\section{Introduction}
The construction of half-maximal-BPS backgrounds with an AdS-factor for the Type II string or for M-theory is a very fertile problem. Illuminated by Maldacena's AdS/CFT correspondence \cite{Maldacena:1997re}, this problem gains significance as the study of non-perturbative aspects of  conformal field theories preserving eight Poincar\'e SUSYs in diverse dimensions. By now, there is a beautiful correspondence between infinite families of AdS$_D\times S^2\times \Sigma_{8-D}$ solutions preserving eight Poincar\'e SUSYs ($D=2,....,7$) and SCFTs in dimension $d=D-1$ with $SU(2)$ R-symmetry.

In fact,  for the case $D=2,3,4,5,6,7$ the formalism, backgrounds and dual field theories are  respectively described in the papers 
\cite{Lozano:2020txg}-\cite{Lozano:2021rmk} (for AdS$_2$),  \cite{Couzens:2017way}-\cite{Lozano:2019ywa}
(for AdS$_3$),  \cite{DHoker:2007hhe}-\cite{Akhond:2021ffz} (for AdS$_4$), \cite{Gaiotto:2009gz}- \cite{Nunez:2018qcj} (for AdS$_5$), 
 \cite{DHoker:2016ujz}-\cite{Legramandi:2021uds} (for AdS$_6$) and  \cite{Apruzzi:2013yva}-\cite{Bergman:2020bvi}  (for AdS$_7$).

A natural next step is to achieve the same classification of Type II (or M-theory) backgrounds and CFTs in situations with less SUSY, typically breaking also part of the  $SU(2)$ R-symmetry. Establishing the correspondence between supergravity backgrounds and precise QFT/CFTs  (in the less SUSY cases) is a very interesting and demanding problem. Few papers have attempted this, mostly due to the technical difficulties in solving BPS equations (generically, non-linear and coupled PDEs). The achievements are  less spectacular than those in  the more symmetric circumstances. See \cite{Bah:2017wxp}-\cite{Bah:2018lyv}, for some works in this direction.

On the field theory side, a popular way of constructing SCFTs
is to consider twisted compactifications of the $d=6~(0,2)$ SCFT on different manifolds. The $(0,2)$ theory then acts as a `mother' of the lower dimensional ones (for example, the class ${\cal S}$ theories).

 In five dimensions there is an infinite family of SCFTs with eight Poincar\'e SUSYs. Its holographic description is given in the papers  \cite{DHoker:2016ujz}-\cite{Legramandi:2021uds}. It is natural to ask if this family of SCFTs can be compactified on two, three, and four manifolds leading to lower dimensional SCFTs. This problem is ideal to tackle using AdS/CFT techniques.

Indeed, a technical tool used in carving the space of possible string backgrounds dual to SCFTs, is to find solutions in D-dimensional AdS-gauged supergravity and lift them to Type II or M-theory. These solutions are typically of the form AdS$_{D-p}\times \Sigma_p$, where $\Sigma_p$ is a compact space, and contain a host of scalars and forms excited. Similar procedures are used to construct duals to defect conformal field theories, see for example \cite{Dibitetto:2017tve}.

This technique  suggests that the dual field theory is a twisted compactification of a $(D-1)$ dimensional CFT that reaches a strongly-coupled lower dimensional IR fixed point of dimension $(D-p-1)$. In this paper, we are interested in finding the flow to these IR fixed points dual to lower dimensional SCFTs.

More concretely, we consider five-dimensional ${\cal N}=2$ SCFTs (for example, the strongly coupled fixed point of 5d linear quiver field theories)
dual to Type IIB solutions with AdS$_6\times S^2$ factors. Holographically, the RG flow to a lower dimensional SCFT is characterized by a solution in an effective six dimensional gauged supergravity \cite{Romans:1985tw} that interpolates between AdS$_6\to$ AdS$_{d+1}\times \Sigma_{5-d}$. This describes the strongly coupled dynamics of the 5d SCFT compactified on $\Sigma_{5-d}$ that reaches a $d$-dimensional CFT$_d$, with less SUSY than the 5d UV one.
In this work, we describe these RG-flows, study fixed points and calculate quantities characterising both the flows and the fixed points. In the same line, SUSY breaking compactifications can be considered.

The contents of this paper are distributed as follows:
in section  \ref{summaryF4}, we summarise the needed aspects of Romans' six dimensional $F_4$ gauged supergravity. In section \ref{section3} we discuss solutions to this gauged supergravity that preserve some fraction of SUSY and have AdS$_4$, AdS$_3$ and AdS$_2$ spaces. We also study compactifications on a circle with SUSY breaking boundary conditions.

Section \ref{sec:AdS6IIB} carefully describes two formulations of the lift to Type IIB of the solutions presented  in section \ref{section3}. We use both the language of \cite{DHoker:2016ysh} and \cite{Legramandi:2021uds} and show the equivalence of the two formulations. In section \ref{newfamsol} the three new
SUSY preserving families of Type IIB backgrounds with AdS$_4$, AdS$_3$ and AdS$_2$ factors are explicitly written. These infinite families are labelled by a function $V(\sigma,\eta)$ that also encodes the dual five-dimensional ${\cal N}=2$ SCFT undergoing compactification. Some field theoretical aspects  of the lower dimensional SCFT$_d$ are discussed in section \ref{FTDsection}. Among other things the number of degrees of freedom is defined and computed in detail. The holographic central charge is calculated at the fixed points and we also present an analog monotonic observable which characterise the dimensional flows. Interestingly, in terms of the quiver parameters, the holographic central charge of the IR fixed point is proportional to the UV one, confirming the picture advocated in \cite{Bobev:2017uzs}.
We also present a dual to a QCD-like quiver theory with massive matter. Section \ref{conclusions} presents a summary and conclusions of this work. Some appendices with technical details complement the presentation.

\section{Summary of six dimensional Romans $F_4$ supergravity}\label{summaryF4}

In this section we give an account of six dimensional Romans'  $F_4$ gauged supergravity  \cite{Romans:1985tw}. We use slightly different conventions (and a different signature) to those  in \cite{Romans:1985tw},
 in order to profit from the lift to Type IIB presented in \cite{Hong:2018amk}, as we discuss in  section \ref{sec:AdS6IIB}.

The six dimensional Romans' F(4) gauged supergravity  is defined in terms of a real scalar field $X$, a three-form 
\begin{equation}
F_3= dA_2 \, ,\label{eq1}
\end{equation} 
an Abelian gauge field $A_1$ with field strength
\begin{equation}
F_2= dA_1+ \frac{2}{3} \tilde{g} A_2 \, ,\label{eq2}
\end{equation} 
and a non-Abelian SU(2) gauge field $A^i$ with  curvature 
\begin{equation}
F^i= dA^i + \frac12 \epsilon^{ijk} A^j \wedge A^k.
\end{equation} 
The parameter $\tilde{g}$ is a coupling in the 6d theory. The units
associated to the various fields and curvatures are,
\begin{equation}
[A_2]=[A_1]=1, \;\;\;[F_2]=[F_3]= m, \;\;\; [A^i]=m, [F^i]= m^2,\;\;\; [X]=1,\;\; [\tilde{g} ]= m.
\label{units}
\end{equation}
The bosonic part of the lagrangian reads,
\begin{eqnarray}
& & \mathcal{L} = R *_6 1+ 4 \frac{*_6 d X \wedge d X}{X^2}- \tilde{g}^2 \left(\frac29 X^{-6} - \frac83 X^{-2}-2 X^2\right) *_6 \label{eq:lagrangian6}\\
& & \qquad  + \frac12 X^4 *_6 F_3 \wedge F_3  - \frac12 X^{-2} \left(*_6 F_2 \wedge F_2 + \frac1{\tilde{g}^2} *_6 F^i \wedge F^i\right) \nonumber\\
& & \qquad -\frac12 \tilde{A}_2 \wedge \left( d A_1 \wedge d A_1 + \frac23 \tilde{g} d A_1 \wedge A_2 + \frac{4}{27} \tilde{g}^2 A_2 \wedge A_2 + \frac1{\tilde{g}^2} F^i \wedge F^i \right). \nonumber
\end{eqnarray}
The equations of motion are
\begin{eqnarray}
& & d(X^4 *_6 F_3) = \frac12 F_2 \wedge F_2 + \frac{1}{2 \gt^2} F^i \wedge F^i + \frac23 \gt X^{-2} *_6 F_2, \label{eqs6d} \\
& & d(X^{-2} *_6 F_2) = - F_2 \wedge F_3 \\
& & D(X^{-2} *_6 F^i)= - F_3 \wedge F^i \\
& & d(X^{-1} *_6 d X) = \frac14 X^4 *_6 F_3 \wedge F_3  - \frac{X^{-2}}{8} \left(*_6 F_2 \wedge F_2 + \frac1{\tilde{g}^2} *_6 F^i \wedge F^i\right) \nonumber \\[2mm]
& & \qquad \qquad \qquad \quad  - \tilde{g}^2 \left(\frac16 X^{-6}  - \frac23 X^{-2}+ \frac12 X^2\right) *_6 1 , \label{eqs6dlast}
\end{eqnarray}
whee we defined  $D$ as the SU(2) gauge-covariant derivative, defined on a differential for $C^i$ by
\begin{equation}
DC^i= dC^i +\epsilon_{ijk}A^j \wedge C^k \, .\label{eq3}
\end{equation}
Finally, the Einstein's equations are
\begin{eqnarray}
& & R_{\mu \nu} = 4 X^{-2} \partial_\mu X \partial_\nu X+ \gt^2   \left(\frac{1}{18} X^{-6}  - \frac23 X^{-2}- \frac12 X^2\right) g_{\mu \nu} + \frac{X^4}{4} \left(F_{3 \, \mu} \cdot F_{3 \, \nu} - \frac{1}{6} g_{\mu \nu} F_3^2\right) \nonumber \\[2mm]
& & \qquad + \frac{X^{-2}}{2} \left(F_{2 \, \mu} \cdot F_{2 \, \nu} - \frac{1}{8} g_{\mu \nu} F_2^2\right) + \frac{X^{-2}}{2 \gt^2} \left(F^i_{2 \, \mu} \cdot F^i_{2 \, \nu} - \frac{1}{8} g_{\mu \nu} (F_2^i)^2\right), \label{Einstein6d}
\end{eqnarray}
where $F_\mu = \iota_\mu F$ is the contraction along the direction $\partial_\mu$, $F \cdot G = F_{\mu_1 \dots \mu_p} G^{\mu_1 \dots \mu_p}$, and $F^2 = F \cdot F$.

The fermionic part of the Lagrangian and SUSY transformations can be found in \cite{Romans:1985tw}.
The simplest possible solution sets all matter fields to zero $A_2=A_1=A^i=0$ and $X=1$. The metric is that of  AdS$_6$ of radius $R^2=\frac{2}{9 \gt^2}$. This solution preserves eight Poincar\'e supercharges.
Let us  now discuss various solutions of this six dimensional system.
\section{Solutions in six dimensional $F_4$ supergravity}\label{section3}
%
%

In this section we discuss various solutions to $F_4$ gauged supergravity. The fixed point solutions described below are not new, in fact they can be found in \cite{Nunez:2001pt} and \cite{Suh:2018tul}. We present them in detail in our conventions and also display new numerical interpolations between fixed points. In all cases we have checked that all the equations (\ref{eqs6d})-(\ref{Einstein6d}) are satisfied.

These solutions describe spontaneous compactifications, triggered by fluxes, of the AdS$_6$ SUSY vacuum of six dimensional supergravity on the manifolds $H_2$, $H_3$, $H_2\times H_2$. They are discussed in sections \ref{AdS4H2}, \ref{AdS3H3}, \ref{AdS2H2H2} respectively.  These solutions preserve different fractions of the original Poincar\'e SUSYs. Finally, a non-SUSY  solution obtained as a double Wick rotation of an AdS$_6$ black hole is described in section \ref{BHWR}.

We start by writing flows between AdS$_6$ and AdS$_4$ when the supergravity compactifies on a compact hyperbolic two-dimensional space

\subsection{AdS$_6 \to $ AdS$_4\times H_2$}\label{AdS4H2}

\begin{figure}[!]
	\centering
	\includegraphics[scale=0.6]{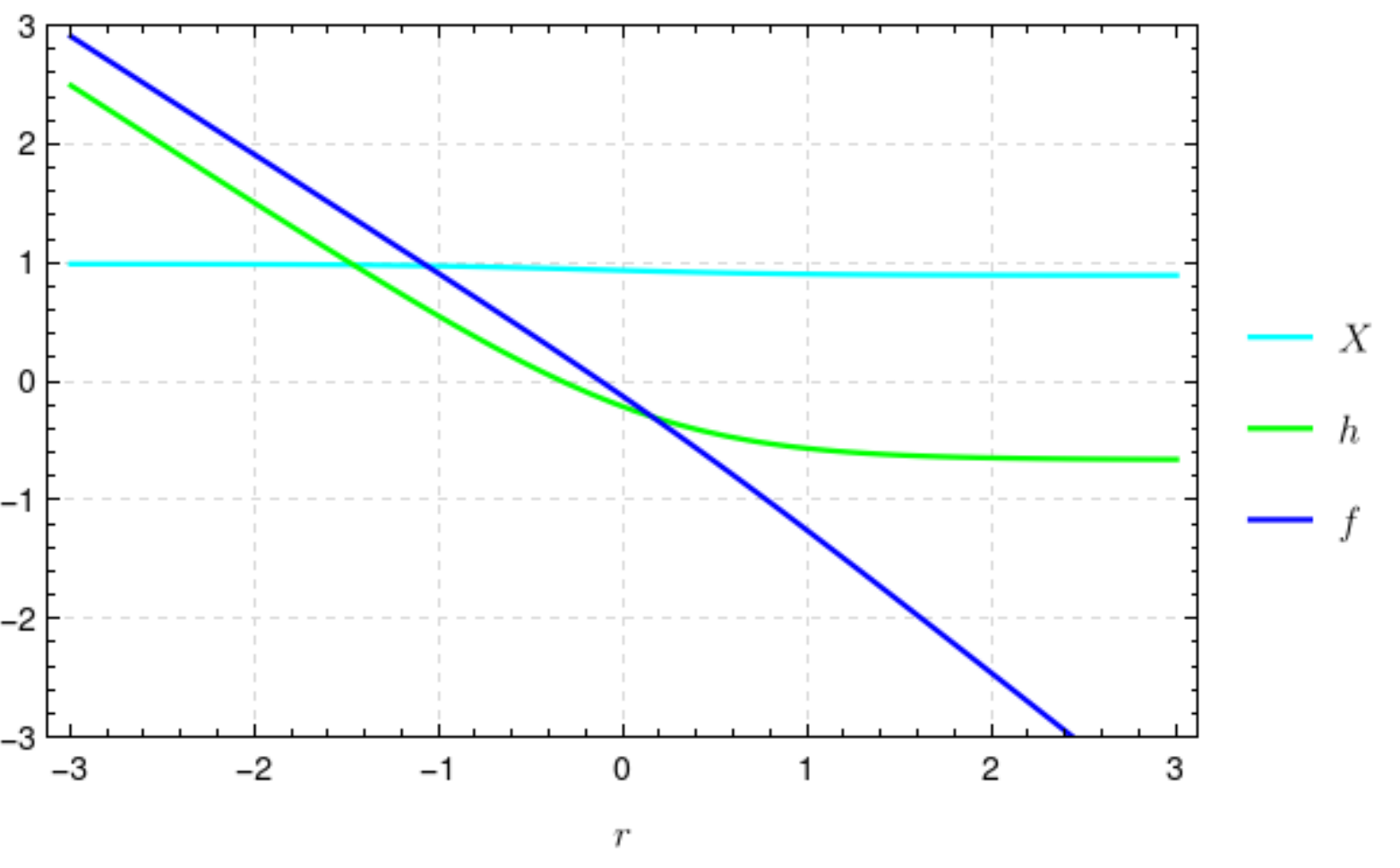}
	\caption{Numerical interpolation between the AdS$_6$ and AdS$_4$ fixed point. The details of the numerical analysis are explained in Appendix \ref{numericsAdS4}. We do not plot $g(r)$ since it is given by $g(r) = -f(r) + \log X(r)$.}
	\label{fig:AdS4}
\end{figure}

This configuration was first discussed in  \cite{Nunez:2001pt}, we review it here in our conventions. We propose a metric
\begin{equation}
\label{eq:AdS4xH2_metric}
d s^2 = e^{2 f(r)}(-dt^2+ dy_1^2+ dy_2^2 + e^{2 g(r)}dr^2) +  e^{2h(r)}\frac{(dz^2+ dx^2)}{z^2} ,
\end{equation}
where $g(r)$ is an arbitrary function which takes into account the reparameterisation invariance of the coordinate $r$. The gauge field configuration breaks SU(2) to U(1)
\begin{equation}
\label{eq:AdS4xH2_Ai}
A^1 = 0 \, , \qquad A^2 = 0 \, , \qquad  A^3 = \frac{1}{z} d x \, .
\end{equation}
The dilaton depends  only on the $r$ coordinate, $X=X(r)$ and the fields $A_1=A_2=0$.  The BPS and the equations of motion are solved if we impose
\begin{eqnarray}
& & X' = -\frac{e^{f+g} \gt }{2\sqrt{2} } \left[X^{-2} -X^2- \frac{1}{2 \gt^2} e^{-2 h } \right] \, ,
\label{martinn1} \\[2mm]
& & h' = -\frac{e^{f+g} \gt }{2\sqrt{2} X } \left[ \frac{X^{-2}}{3} +X^2- \frac{3}{2 \gt^2} e^{-2 h } \right] \, ,
\label{martinn2} \\[2mm] 
& & f' = -\frac{e^{f+g} \gt }{2\sqrt{2} X } \left[ \frac{X^{-2}}{3} +X^2+ \frac{1}{2 \gt^2} e^{-2 h } \right] \, .
\label{martinn3}
\end{eqnarray}
The solutions to these equations generate backgrounds preserving four Poincar\'e SUSYs (${\cal N}=2 $ in the language of the dual 3d SCFT), see  \cite{Nunez:2001pt} for a detailed analysis.
The BPS equations admit a fixed point solution by requiring that both $X$ and $h$ are constants,
\begin{equation}
\label{eq:AdS_4_fixed}
X^4 = \frac23 \, , \qquad e^{2 h} = \sqrt{\frac32} \gt^{-2} \, , \qquad e^{2g} = e^{-2 f} \frac{\sqrt{6}}{\gt^2} (f')^2 \, .
\end{equation}
Plugging this result in eq.\eqref{eq:AdS4xH2_metric}, the metric is  AdS$_4 \times H_2$ (where we can now use $f$ as a coordinate).
Also notice that for small $r$, the expansion
\begin{equation}
f = - \log r + O(r^2) \, , \qquad h = - \log r + O(r^2) \, , \qquad X = 1 + O(r^2) \, , \qquad g = O(r^2) ,\label{oooppp}
\end{equation}
gives a power series solution to the BPS equations which approaches AdS$_6$ as $r \to 0$\footnote{ Choosing $\tilde{g} = 3 / \sqrt{2}$  gives the metric on AdS$_6$ with a unit radius. We can choose to work with a generic $\gt$ by adding a constant factor in the expansion of $f$ modifying the AdS$_6$ radius.}.

The solution which interpolates among these two fixed points is shown in Figure \ref{fig:AdS4}. For the numerical analysis we have set $\tilde{g} = 3 / \sqrt{2}$ and $g(r) = -f(r) + \log X(r)$.
With this choice of radial coordinate---different from the one in eq.(\ref{oooppp}), we find the AdS$_4$ fixed point at $r\to\infty$ and the AdS$_6$ one at $r\to -\infty$.
Indeed, at the AdS$_4$ fixed point we have
\begin{equation}
r \gg 0 \qquad X = \sqrt[4]{\frac23} = 0.90 \, , \qquad  h = \frac14 \log \frac{2}{3^3} = -0.65 \, , \qquad f= -\sqrt{\frac{3}{2}} r
\end{equation}
consistently with eq.\eqref{eq:AdS_4_fixed}. In the AdS$_6$ fixed point we have
\begin{equation}
r \ll 0 \qquad  X = 1  \, , \qquad  h = -r \, , \qquad f= - r
\end{equation}
which for $r\to -\infty$ approaches the metric of AdS$_6$. See Figure \ref{fig:AdS4} for a plot of the numerical solutions. Details of the numerical analysis can be found in Appendix \ref{numericsAdS4}.
Let us now discuss spontaneous compactifications to AdS$_3$.

\subsection{AdS$_6 \to $ AdS$_3\times H_3$}\label{AdS3H3}
In this section we discuss a flow from AdS$_6$ to a solution of the form AdS$_3\times H_3$, first found in
\cite{Nunez:2001pt}.
Interestingly,  this background needs the non-Abelian character of the gauge field $A^{i}$. The six dimensional configuration reads
\begin{eqnarray}
& & ds^2= e^{2 f(r)}(-dt^2 + dy^2+ e^{2g(r)} dr^2)+ e^{2 h(r)}\frac{(dx_1^2+ dx_2^2+dz^2)}{z^2},\label{configH3}\\
& & A^1= -\frac{1}{z} d x_1 \, , \qquad A^2 = 0 \, , \qquad A^3 = -  \frac{1}{z} d x_2~, \;\;\;\;\; X=X(r) \, .\nonumber
\end{eqnarray}
All the other fields are set to zero.
The BPS equations are
\begin{eqnarray}
& & X'= \frac{\tilde{g} e^{f+g}}{4\sqrt{2}} \left(2 X^2 - 2 X^{-2} +\frac{3}{\tilde{g}^2}e^{-2h} \right) \, ,\label{BPSH3X}\\
& & f'= -\frac{\tilde{g} e^{f+g}}{4\sqrt{2} X} \left(2 X^2 +\frac{2 X^{-2} }{3} +\frac{3}{\tilde{g}^2}e^{-2h} \right) \, ,\label{BPSH3f}\\
& & h'=- \frac{\tilde{g} e^{f+g}}{4\sqrt{2} X} \left(2 X^2 +\frac{2 X^{-2} }{3} -\frac{5}{\tilde{g}^2}e^{-2h} \right) \, .\label{BPSH3h}
\end{eqnarray}
Any solution of the BPS equations (\ref{BPSH3X})-(\ref{BPSH3h}) is a solution of the equations of motion (\ref{eqs6d})-(\ref{Einstein6d}). These backgrounds will preserve two Poincar\'e SUSYs, as analysed in  \cite{Nunez:2001pt}.

\begin{figure}[!]
	\centering
	\includegraphics[scale=0.6]{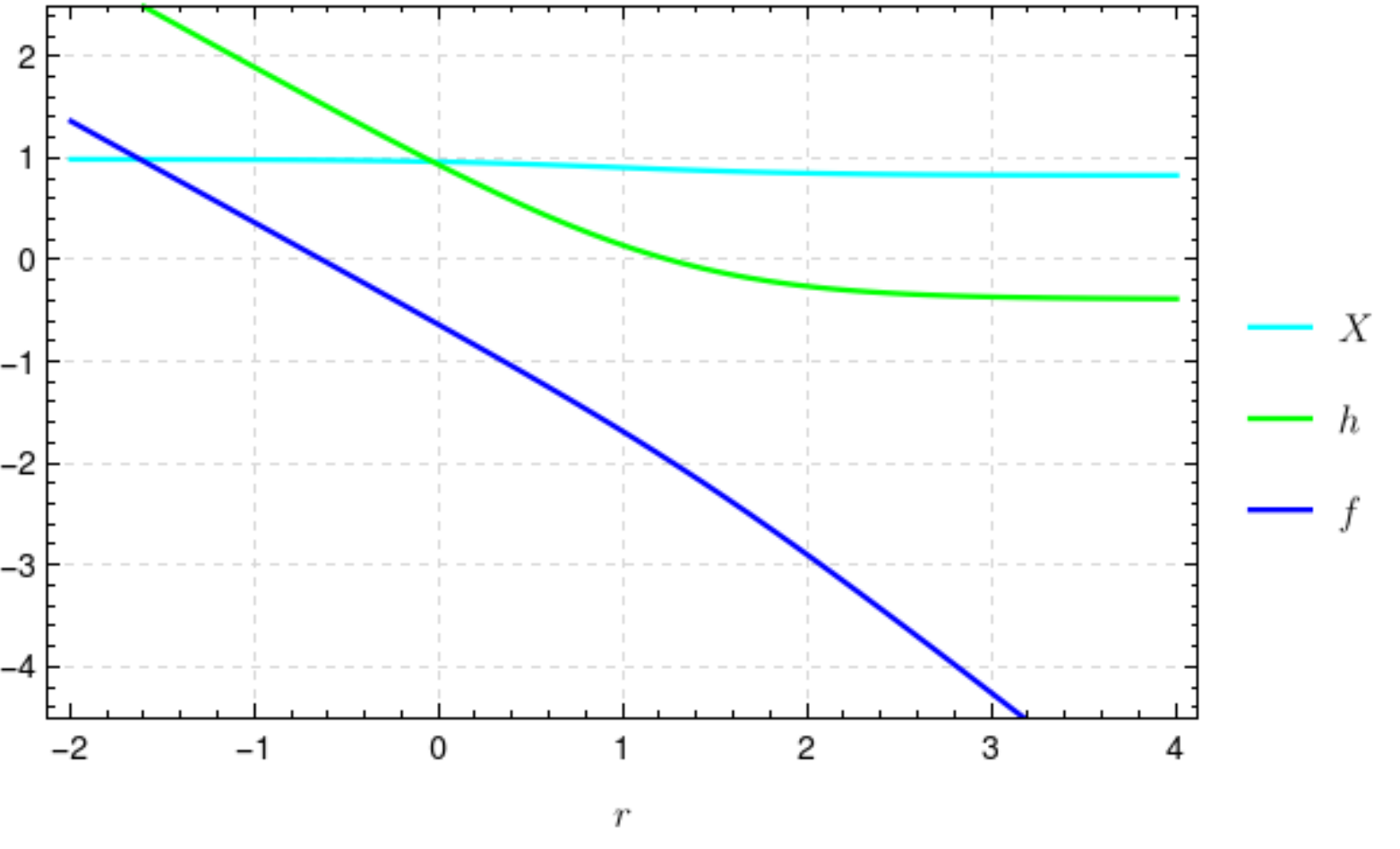}
	\caption{Numerical interpolation between the AdS$_6$ and AdS$_3$ fixed point. The details of the numerical analysis are explained in appendix \ref{numericsAdS3}. Again, $g$ is given by $g = -f + \log X$. }
	\label{fig:AdS3}
\end{figure}

One  possible solution is $X=1 + O(r^2)$ for small $r$ and $f\sim h\sim -\log r$. This reproduces AdS$_6$ (for $r\to 0$). There is also a fixed point solution to the equations,
\begin{equation}
X^4=\frac{1}{2} \, , \qquad  e^{2h}= \frac{3}{\sqrt{2} \tilde{g}^2} \, , \qquad  e^{2g}= e^{-2 f} \frac{9}{2^{5/2}}  \frac{1}{\tilde{g}^2} (f')^2 \, ,\label{AdS3fixedpoint}
\end{equation}
leading to a  space of the form AdS$_3\times H_3$, with a fixed size $H_3$.

The solution describing the flow AdS$_6\to $ AdS$_3\times H_3$ is plotted in Figure \ref{fig:AdS3}. The details of the numerical analysis are written in Appendix \ref{numericsAdS3}. 
As above, we set $\tilde{g} = 3 / \sqrt{2}$ and we use a convenient radial coordinate  set by the choice $g(r) = -f(r) + \log X(r)$. Using this radial coordinate the AdS$_6$ fixed point is found as $r\to-\infty$, while the AdS$_3$ fixed point is found at $r\to+\infty$. Indeed,  we  have
\begin{equation}
r \gg 0 \qquad X = \sqrt[4]{\frac12} = 0.84 \, , \qquad  h = \frac14 \log \frac{2}{9} = -0.38 \, , \qquad f= -\sqrt{2} r
\end{equation}
which is consistent with eq.\eqref{AdS3fixedpoint}. In the AdS$_6$  fixed point we have
\begin{equation}
r \ll 0 \qquad  X = 1 \, , \qquad  h = -r \, , \qquad f= - r~.
\end{equation}
Let us now move to analyse spontaneous compactifications to AdS$_2$.

\subsection{AdS$_6 \to $ AdS$_2\times H_2^{(1)}\times H_2^{(2)}$}\label{AdS2H2H2}

The backgrounds we summarise in this section were first discussed in \cite{Suh:2018tul}.
As the solution in section \ref{AdS4H2}, he configuration breaks the SU(2) gauge symmetry to U(1), but in this case we also have $A_2$ turned on
\begin{eqnarray}
& & d s^2_6 = e^{2 f(r)}(-dt^2+ e^{2 g(r)}dr^2) +  e^{2h_1(r)}\frac{(dz_1^2+ dx_1^2)}{z_1^2} +  e^{2h_2(r)}\frac{(dz_2^2+ dx_2^2)}{z_2^2} , \\
& & 
A^1 = 0 \, , \qquad A^2 = 0 \, , \qquad  A^3 = \frac{1}{z_1} d x_1+ \frac{1}{z_2} d x_2 \, , \qquad  X=X(r)\label{eq:AdS2xH2XH2_Ai}\\
& & 
A_3=0 \, , \qquad A_2 = \frac{9}{4 \gt^4} X(r)^2 e^{2(f-h_1-h_2) + g} d t \wedge d r \, .\label{A2}
\end{eqnarray}
With these definitions and using eqs.(\ref{eq1})-(\ref{eq2}), we have 
\begin{equation}
F_3 = d A_2 = 0 \, , \qquad F_2 = \frac{2}{3} \gt A_2 \neq 0 .
\end{equation}
The amount of SUSY preserved and the BPS system for this configuration were studied in  \cite{Suh:2018tul}. The BPS equations read,
\begin{eqnarray}
& & h_1'=-\frac{\gt e^{f+g}}{2 \sqrt{2} X} \left(X^2+\frac{X^{-2}}{3}-\frac{1}{2 \gt^2} (3 e^{-2 h_1}-e^{-2 h_2})-\frac{3 X^2}{4 \gt^4}  e^{-2 h_1-2 h_2}\right) \, , \\
& & h_2'=-\frac{\gt e^{f+g}}{2 \sqrt{2} X} \left(X^2+\frac{X^{-2}}{3} -\frac{1}{2 \gt^2} (3 e^{-2 h_2}-e^{-2 h_1})-\frac{3 X^2}{4 \gt^4}  e^{-2 h_1-2 h_2}\right) \, ,  \\
& &  f'=-\frac{\gt e^{f+g}}{2 \sqrt{2} X} \left(X^2+\frac{X^{-2}}{3} +\frac{1}{2 \gt^2} (e^{-2 h_1}+e^{-2 h_2})+\frac{9 X^2}{4 \gt^4}  e^{-2 h_1-2 h_2}\right) \, , \\
& &  X'=-\frac{\gt e^{f+g}}{2 \sqrt{2}} \left(-X^2+ X^{-2}-\frac{1}{2 \gt^2} (e^{-2 h_1}+e^{-2 h_2})+\frac{3 X^2}{4 \gt^4}  e^{-2 h_1-2 h_2}\right) \, .
\end{eqnarray}
The system admits a fixed-point solution
\begin{equation}
e^{h_1} = e^{h_2} = \sqrt[4]{\frac{3}{2}} \gt^{-1} \, , \qquad X = \sqrt[4]{\frac{2}{3}} \, , \qquad e^{2f + 2g} = \sqrt{\frac{3}{2}} \frac{(f')^2}{2 \gt^2} \, ,\label{AdS6toAdS2fixed}
\end{equation}
which correspond to an AdS$_2 \times H_2 \times H_2$ metric. The BPS system can also be solved with power series expansion around $r\sim 0$ whose leading terms are
\begin{equation}
f (r)= - \log r \, , \qquad h_1(r) = - \log r \, , \qquad h_2(r) = - \log r \, , \qquad X (r)= 1 \, , \qquad g(r) = 0 
\end{equation}
which approaches AdS$_6$ as $r \to 0$. 

\begin{figure}[!]
	\centering
	\includegraphics[scale=0.6]{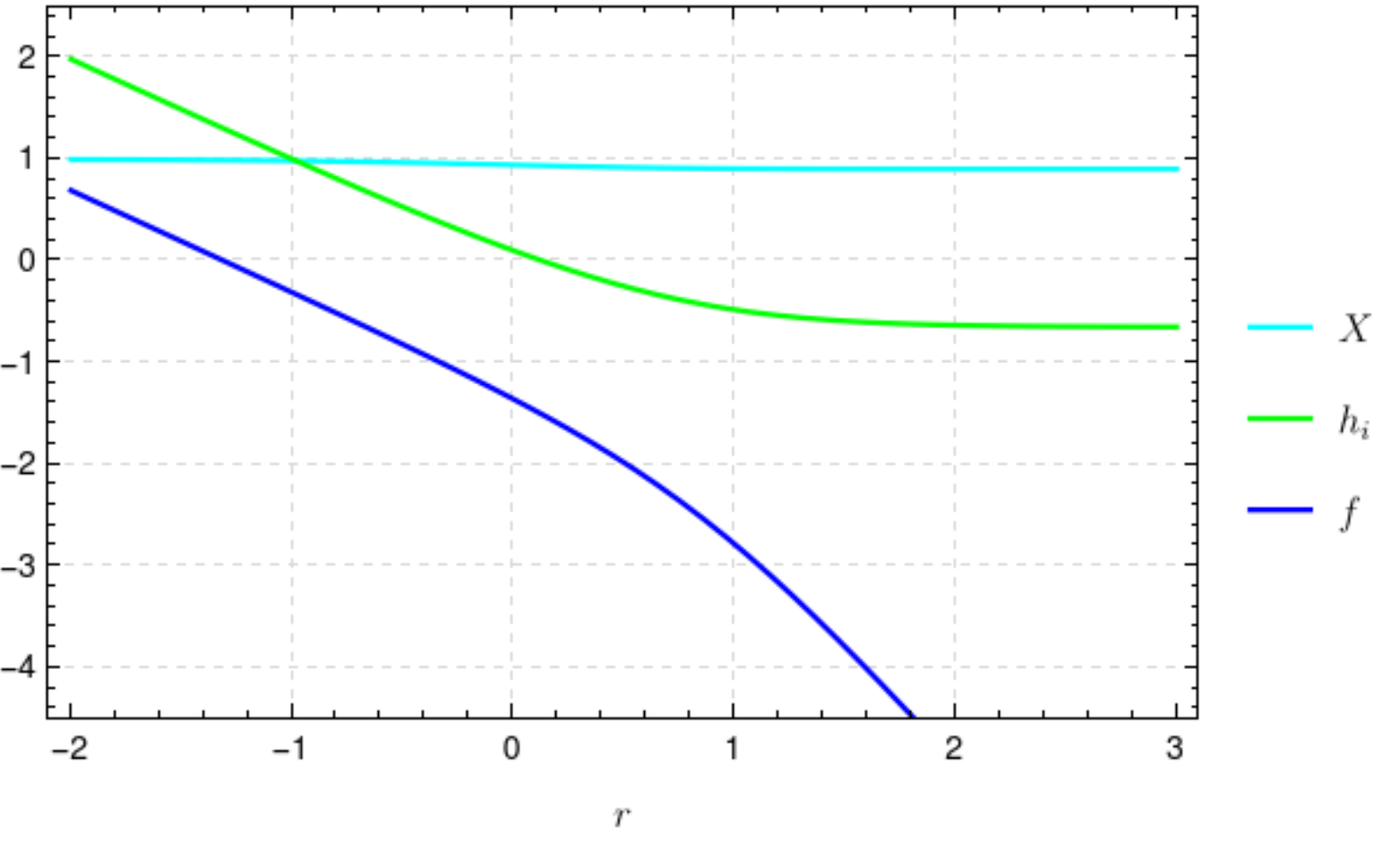}
	\caption{Numerical interpolation between the AdS$_6$ and AdS$_2$ fixed point. The details of the numerical analysis are explained in appendix \ref{numericsAdS2}. Again, $g$ is given by $g = -f + \log X$ and we have assumed $h_1 = h_2$ along the flow.}
	\label{fig:AdS2}
\end{figure}

Since $h_1 = h_2$ at both  fixed points, we apply the simplifying ansatz $h_1 = h_2$ along the full AdS$_6\to $ AdS$_2\times H_2 \times H_2$ flow. Notice that this choice makes one equation in the BPS system redundant.
The flow is plotted in Figure \ref{fig:AdS2} and we refer to appendix \ref{numericsAdS2} for the details about the numerical analysis. 
Again we set $\tilde{g} = 3 / \sqrt{2}$ and $g(r) = -f(r) + \log X(r)$;
the AdS$_2$ fixed point is found at $r\to+\infty$:
\begin{equation}
r \gg 0 \qquad X = \sqrt[4]{\frac23} = 0.90 \, , \qquad  h_i = \frac14 \log \frac{2}{27} = -0.65 \, , \qquad f= -\sqrt{6} r
\end{equation}
which is consistent with eq.\eqref{AdS6toAdS2fixed},while we obtain the AdS$_6$ for
\begin{equation}
r \ll 0 \qquad  X = 1 \, , \qquad  h_i = -r \, , \qquad f= - r~.
\end{equation}

Finally, let us discuss a solution describing the compactification on a circle.
\subsection{AdS$_6 \to R^{1,3}\times S^1$}\label{BHWR}
In this section we discuss  a very simple solution. It can be obtained by Wick rotating a non supersymmetric black hole in AdS$_6$.
In fact, consider the solution with all the fields turned off  $A_2=A_1=A^i=0$ and $X=1$. The metric is,
\begin{equation}
ds^2 = e^{2\rho}\left(- d\tau^2+ d\vec{x}_3^2  \right) +\frac{d\rho^2}{g(\rho)}+ e^{2\rho} g(\rho) d\psi^2.
\end{equation}
If $g(\rho) = 1$ this is the supersymmetric AdS$_6$ solution mentioned below eq.(\ref{Einstein6d}). Here we set
\begin{eqnarray}
g(\rho)=  1-e^{5 (\rho_* -\rho)}.
\end{eqnarray}
The  equations of motion (\ref{eqs6d})-(\ref{Einstein6d}) are  satisfied. Supersymmetry is broken, unless $\rho_*\to -\infty$.
The variable $\rho$, ranges in $[\rho_*,\infty)$. For this background to be smooth we need to set up constants such that a conical singularity is avoided. This fixes 
$\rho_*= \log(\frac{2}{5\sqrt{5}})$. In fact, close to the end of the space,   $x= (\rho-\rho_*)\to 0$,
\begin{eqnarray}
\frac{d\rho^2}{g(\rho)}+ e^{2\rho} g(\rho) d\psi^2\approx \frac{dx^2}{5x} + 5 x e^{2\rho_*} d\psi^2= du^2+ \frac{125e^{2\rho_*}}{4} u^2 d\psi^2= du^2+ u^2 d\psi^2.\nonumber
\end{eqnarray}
In the next section, we consider all the six-dimensional backgrounds presented in this section and lift them to Type IIB. We will use the lifting formulas of \cite{Hong:2018amk}, written in the formulation of \cite{Legramandi:2021uds}, \cite{Apruzzi:2018cvq}. This will provide {\it new} supersymmetric
Type IIB families of backgrounds with AdS$_4$, AdS$_3$, and AdS$_2$ factors. After lifting, the solution in section \ref{BHWR} provides a non-SUSY background interpolating between AdS$_6$ and $R^{1,3}\times S^1$ in Type IIB. After that, in section \ref{FTDsection}, we present a dual CFT/QFT interpretation of these backgrounds and compute interesting observables. 

Before all of this, we carefully  describe the uplifting formulas and the dictionary between the formalism of \cite{Hong:2018amk} and that of \cite{Legramandi:2021uds}, \cite{Apruzzi:2018cvq}. To this we turn.


\section{Type IIB  AdS$_6$ background}\label{sec:AdS6IIB}
%
%
%
In this section we present two different but equivalent ways of writing an infinite family  of AdS$_6$ background in type IIB supergravity. These backgrounds preserve eight Poincar\'e supercharges.  
Aside from the $SO(2,5) $ isometry realised by the AdS$_6$ factor, the $SU(2)$ isometries of a two-sphere are also imposed. The internal space has just two  unconstrained dimensions.

It is convenient to parameterise the two-dimensional space using complex coordinates. This leads to expressing the background in terms of two holomorphic functions. 
This infinite family of solutions was first written in \cite{DHoker:2016ujz}. In this paper, we use the notation of \cite{Hong:2018amk}. 

A second possibility, first analysed in \cite{Apruzzi:2018cvq} and further developed in 
\cite{Legramandi:2021uds},
 is to express the background  in terms of a real function which solves a Laplace equation. Bellow we describe both formulations and show their equivalence.

\subsection{Using holomorphic functions}
\label{sub:holo_func}

This first formulation applies in the case for which  all the fields in the background are locally written in terms of two unconstrained holomorphic functions  $\mathcal{A}_\pm(z)$ of a complex coordinate $z$. 
To succinctly describe the formalism is convenient to define:
\begin{itemize}
\item{An auxiliary holomorphic function $\mathcal{B}$ in terms of the  differential equation,
\begin{equation}
\partial_z \mathcal{B} = \mathcal{A}_+\partial_z\mathcal{A}_- - \mathcal{A}_-\partial_z\mathcal{A}_+ \, .\label{Bz}
\end{equation}
}
\item{
The real functions,
\begin{equation}
\begin{split}
&\mathcal{G} = |\mathcal{A}_+|^2-|\mathcal{A}_-|^2+2 \text{Re} \mathcal{B} \, , \qquad \kappa^2 = - \partial_z \partial_{\bar{z}} \mathcal{G} = |\partial_z\mathcal{A}_-|^2-|\partial_z\mathcal{A}_+|^2 \, , \\
&\mathcal{Y} = \frac{\kappa^2 \mathcal{G}}{|\partial_z \mathcal{G}|^2} \, , \qquad \qquad \qquad \qquad \, \, \, \, \mathcal{D} = 1 + \frac{2}{3 \mathcal{Y}} \, .
\end{split}
\end{equation}
}
\end{itemize}
Using this, 
the string-frame metric reads
\begin{equation}
d s^2_{st} = e^{\frac{\Phi}{2}}\left(f_1  d s^2(\text{AdS}_6) + f_2 d s^2(S^2)+ 4 f_3 d z d \bar{z}  \right),
\end{equation}
where
\begin{equation}
f_1 = \frac{\kappa^2 \sqrt{\mathcal{D}}}{f_3}  \, , \qquad f_2 = \frac{1}{9 f_3} \frac{\kappa^2}{\sqrt{\mathcal{D}}} \, , \qquad f_3^2 =  \frac{\kappa^4 \sqrt{\mathcal{D}}}{6 \mathcal{G}} \, .
\end{equation}
The rest of the Type IIB  fields,  are expressed using the  $SU(1,1)$ covariant formalism, 
\begin{eqnarray}
& & \frac{1 + i \tau}{1-i \tau} = \frac{\mathcal{A}_+ - \bar{\mathcal{A}}_- - \mathcal{C} / \sqrt{\mathcal{D}}}{\bar{\mathcal{A}}_+ - \mathcal{A}_- + \bar{\mathcal{C}} / \sqrt{\mathcal{D}}} \, , \qquad \tau= C_0 + i e^{-\Phi} \, , \nonumber\\[2mm]
& & B_2 +i C_2 = \frac{2 i}{9} \left[\frac{\mathcal{C}}{\mathcal{D}} - 3 (\mathcal{A}_+ + \mathcal{A}_- )\right] \text{Vol}(S^2) \, , \nonumber \\[2mm]
& & F_5 = 0 \, . 
\end{eqnarray}
We have also defined 
\begin{equation}
\mathcal{C} = \frac{\partial_z \mathcal{A}_+ \partial_{\bar{z}} \mathcal{G}+\partial_{\bar{z}} \bar{\mathcal{A}}_- \partial_z \mathcal{G}}{\kappa^2} \, .\label{Cz}
\end{equation}
In the expressions above, $C_0$ and $C_2$ are the RR potential for $F_1$ and $F_3$ respectively while $B_2$ is the potential for the NSNS three form $H$. Respect to \cite{Hong:2018amk}, we have set $c_6 = 1$.
In summary, the backgrounds of eqs.(\ref{Bz})-(\ref{Cz}) represent an infinite family of AdS$_6\times S^2$ solutions of type IIB preserving eight Poincar\'e SUSYs.
 $\mathcal{A}_\pm(z)$.
 
Next, we present the formulation in terms of one real function \cite{Apruzzi:2018cvq},
\cite{Legramandi:2021uds},
 and prove the equivalence of this second formulation with the one of  \cite{DHoker:2016ujz}, above described.
\subsection{Using a real potential}
\label{sub:real_func}

Another formulation of AdS$_6$ backgrounds with $SU(2)$ isometry,  preserving eight supercharges was given in \cite{Apruzzi:2018cvq}. All the fields depend on a real potential $V$ which has support on the two-dimensional internal space. We will sketch the relation between this infinite family of  backgrounds and the one summarised  above in the next section.

It is convenient to parameterise the Riemann surface in terms of two real coordinates, $(\sigma, \eta)$, and use a `potential function' $V(\sigma,\eta)$  that solves the  Laplace partial differential equation
\begin{equation}
\partial_\sigma \left(\sigma^2 \partial_\sigma V\right) +\sigma^2 \partial^2_\eta V=0 \, .\label{diffeq} 
\end{equation}
The type IIB background metric in string frame is
\begin{eqnarray}
& & ds_{10,st}^2= \tilde{f}_1 \Big[ds^2(\text{AdS}_6) + \tilde{f}_2 d s^2 (S^2)+ \tilde{f}_3 (d\sigma^2+d\eta^2) \Big] \, , \quad \tilde{f}_1= \frac{2}{3}\sqrt{\sigma^2 +\frac{3\sigma \partial_\sigma V}{\partial^2_\eta V}}, \nonumber\\[2mm]
& &\tilde{f}_2= \frac{\partial_\sigma V \partial^2_\eta V}{3\tilde{\Lambda}}, \quad \tilde{f}_3= \frac{\partial^2_\eta V}{3\sigma \partial_\sigma V}, \quad \tilde{\Lambda}=3 \partial _{\eta}^2 V \partial _{\sigma }V+ \sigma \left[\left(\partial^2_{\eta \sigma} V \right)^2+\left(\partial _{\eta}^2 V\right)^2\right] \label{background1} 
\end{eqnarray}
The fluxes are given by the following expressions,
\begin{eqnarray}
& & B_2= \tilde{f}_4 \text{Vol}(S^2)=\frac{2}{9}\left(\eta -\frac{(\sigma \partial_\sigma V) (\partial_\sigma\partial_\eta V)}{\tilde{\Lambda}} \right) \text{Vol}(S^2) \, , \nonumber \\
& & C_2= \tilde{f}_5 \text{Vol}(S^2)=4\left( V- \frac{\sigma\partial_\sigma V}{\tilde{\Lambda}} (\partial_\eta V (\partial_\sigma \partial_\eta V) -3 (\partial^2_\eta V)(\partial_\sigma V)) \right) \text{Vol}(S^2) \, , \nonumber\\ [2mm]
& & e^{-2\Phi} =\tilde{f}_6=18^2 \frac{3\sigma^2 \partial_\sigma V \partial^2_\eta V}{(3 \partial_\sigma V +\sigma \partial^2_\eta V)^2}\tilde{\Lambda} \, , \qquad C_0=\tilde{f}_7=18\left( \partial_\eta V + \frac{(3\sigma \partial_\sigma V) (\partial_\sigma\partial_\eta V )}{3\partial_\sigma V +\sigma \partial^2_\eta V}  \right) \, ,\nonumber \\
& & F_5 = 0 \, . \label{background2}
\end{eqnarray}
It was shown in \cite{Legramandi:2021uds} that, subject to eq.(\ref{diffeq}), all the equations of motion of the configuration in eqs.(\ref{background1})-(\ref{background2}) are satisfied, while supersymmetry is discussed in \cite{Apruzzi:2018cvq}.
Let us now explain the map between these two infinite family of solutions.
\subsection{Matching the backgrounds}
As explained in \cite{Legramandi:2021uds}, the two  families of solutions discussed in the sections  \ref{sub:holo_func} and \ref{sub:real_func} actually describe the same configuration.
 In fact, we  can define a complex coordinate from the two real ones $(\sigma, \eta)$
\begin{equation}
z = \sigma - i \eta \, .
\end{equation}
In terms of this complex coordinate, we choose
\begin{equation}
\label{eq:A1A2def}
\mathcal{A}_\pm = \mp \frac{z}{6} - 6\partial_z (\sigma V) \, , 
\end{equation}
and we recover the family in eqs.\eqref{background1}-\eqref{background2} from the backgrounds in eqs.(\ref{Bz})-(\ref{Cz}).

Let us further  discuss the identification in eq. \eqref{eq:A1A2def}. Notice that defining $ V = \hat{V} /\sigma$,  eq.\eqref{diffeq} becomes
\begin{equation}
(\partial^2_\sigma + \partial^2_\eta) \hat{V} = 0 \, .
\end{equation}
Therefore $\hat{V}$ is a  real harmonic function. Hence it defines one holomorphic function $\mathcal{V}(z)$
\begin{equation}
\hat{V}= \mathcal{V}(z) + \overline{\mathcal{V}(z)} \, .
\end{equation}
It might seem that  the backgrounds in section \ref{sub:real_func} are less general than those in section \ref{sub:holo_func}.  Whilst the backgrounds in eqs.(\ref{Bz})-(\ref{Cz}) needed two holomorphic functions  $\mathcal{A}_\pm (z)$, those in eqs. \eqref{background1}-\eqref{background2}  can be defined with just one holomorphic function $ \mathcal{V}(z)$.
But,
notice that the backgrounds of section \ref{sub:holo_func} are  invariant under conformal transformations $z \mapsto \mathcal{F}(z)$, which means that one of the holomorphic functions  $\mathcal{A}_\pm$ can be gauged away. 

We can use eq.\eqref{eq:A1A2def} to make more explicit the dictionary between sections \ref{sub:holo_func} and \ref{sub:real_func}. We have
\begin{eqnarray}
& &\mathcal{G} = 4\sigma^2 \partial_{\sigma} V \, ,
\qquad \kappa^2 = 2 \sigma \partial_\eta^2 V \, , \qquad \mathcal{Y} = \frac{2 \partial_\sigma V\partial _{\eta}^2 V}{\tilde\Lambda-3\partial _{\eta}^2 V\partial _{\sigma }V} \, , \nonumber \\[2mm]
& &\mathcal{C} = -\frac{\sigma \partial_{\eta \sigma}^2 V(18 \partial_{\eta \sigma}^2 (\sigma V)-i)}{\partial_{\eta}^2 V}-6 \sigma^2 \partial_{\eta}^2 V \label{eq:dictionary}
\end{eqnarray}

Equations \eqref{eq:A1A2def} and \eqref{eq:dictionary} can be used to write the uplift of a generic solution in Romans' supergravity to type IIB in terms of the potential $V$. Below, we explain this.
\section{The new families of solutions}\label{newfamsol}


In this section we show how   to write the uplift of  any configuration of Romans supergravity to type IIB.
Then, we explicitly write three new infinite families of SUSY type IIB backgrounds with AdS$_4$, AdS$_3$ and AdS$_2$ factors. We finally discuss the SUSY breaking  compactification AdS$_6\to R^{1,3}\times S^1$.

 The lift  to type IIB of generic configurations of Romans' supergravity was first derived in \cite{Hong:2018amk}\footnote{A lift to Massive IIA was presented in \cite{Cvetic:1999un}.}. 
We use eqs.\eqref{eq:A1A2def} and \eqref{eq:dictionary} to rewrite the result in \cite{Hong:2018amk} in terms of the potential $V(\sigma,\eta)$. Whilst we rely on \cite{Hong:2018amk} for the general uplift, we have explicitly checked that the type IIB equation of motions are solved for all the cases presented in section \ref{section3}. 

A generic solution in Romans' $F_4$ gauged supergravity lifts to a configuration in type IIB  given by,
\begin{eqnarray}
& & d s^2_{st} = f_1 \left( d s^2_6 + f_2 d s^2 (\tilde{S}^2)+ f_3 d s^2 (R^2) \right) \, \label{tendconfig}\\[2mm]
& &C_0 = f_7,\;\;\;\;  e^{-2\Phi} = f_6, \qquad F_5 = 4 (G_5 +*_{10} G_5), \nonumber\\[2mm]
& &B_2 = f_4\text{Vol}(\tilde{S}^2)- \frac{2 \gt}{9}  \sigma F_2  - \frac{2}{9} \eta y^i F^i \, , \nonumber\\[2mm]
& & C_2 = f_5\text{Vol}(\tilde{S}^2)- 4 \gt  \sigma \partial_\eta V F_2-4  \partial_\sigma (\sigma V) y^i F^i \, , \nonumber
\end{eqnarray}
where $d s^2_6$ is defined in terms of the gauged-supergravity metric as 
\begin{equation}
d s^2_6=  \frac{2\tilde{g}^2}{9}  d s^2_{\text{gauged sugra.}}
\end{equation}
The functions $f_i$ are a deformation of the $\tilde{f}_i$ defined in eqs. \eqref{background1}-\eqref{background2}:
\begin{eqnarray}
& & f_1 =\frac{2}{3 X^2} \left( \sigma^2 + \frac{3 X^4\sigma \partial_{\sigma}V}{\partial_{\eta}^2 V}\right)^{1/2}, \qquad f_2 = \frac{X^2\partial_{\sigma}V\partial_{\eta}^2V}{3{\Lambda}}, \qquad  f_3 = \frac{X^2\partial_{\eta}^2V}{3\sigma\partial_{\sigma}V} \label{defi} \\
& &   f_6= (18)^2 \frac{{3 X^4 (\sigma^2 \partial_\sigma V) (\partial_{\eta}^2 V) }}{\left(3 X^4 \partial_\sigma V+\sigma  \partial_{\eta}^2 V\right)^2} \Lambda , \qquad f_7= 18 \left(\partial_\eta V+\frac{3 X^4 \sigma  \partial_\sigma V \partial_{\sigma \eta}^2V}{3 X^4 \partial_\sigma V+\sigma  \partial_{\eta}^2 V}\right), \nonumber\\
& & f_4= \frac{2}{9} \left(\eta -\frac{\sigma  \partial_{\sigma} V \partial_{\sigma \eta}^2 V}{\Lambda }\right),\qquad f_5=4 \left(V-\frac{\sigma   \partial_{\sigma} V \left( \partial_{\eta} V \partial_{\sigma \eta}^2 V-3 X^4 \partial_{\eta}^2 V \partial_{\sigma} V \right)}{\Lambda}\right),\nonumber\\
& & \Lambda=3 X^4 \partial _{\eta}^2 V \partial _{\sigma }V+ \sigma \left[\left(\partial^2_{\eta \sigma} V \right)^2+\left(\partial _{\eta}^2 V\right)^2\right] ,\nonumber
\end{eqnarray}
and reduce to them for $X=1$.
The sphere $\tilde{S}^2$ is fibered over the six-dimensional spacetime, 
\begin{equation}
\text{Vol}(\tilde{S}^2) =\epsilon^{ijk} y^iDy^j\wedge Dy^k , \qquad d s^2_{\tilde{S}^2}= Dy^i Dy^i .\label{eq4}
\end{equation}
Here, $y^i$ are the embedding coordinate of the $S^2$ which can be chosen as
\begin{equation}
\label{eq:S2_embedding}
y^1= \sin\theta\sin\varphi, \qquad y^2= \sin\theta \cos\varphi, \qquad y^3= -\cos\theta , \\
\end{equation}
The symbol $D$ denotes the covariant derivative, as defined in eq.(\ref{eq3}). Finally, 
$G_5$ is a differential form defined as
\begin{equation}
\label{eq:G5_def}
G_5 = - \frac{4}{9} \gt^2 \sigma X^4  \partial_\eta^2 V *_6 F_3 \wedge d \eta \wedge d \sigma+ \frac{2}{3X^2} \gt (*_6 F_2) \wedge *_2 d (\sigma ^2 \partial_\sigma V)-\frac{2}{3 X^2} (*_6 F^i) \wedge D (y^i \sigma ^2 \partial_\sigma V) \, .
\end{equation}

Let us make a quick summary. Any six dimensional configuration solution of the equation of motion (\ref{eqs6d})-(\ref{Einstein6d}) can be lifted to  type IIB as in eqs.(\ref{tendconfig})-(\ref{eq:G5_def}) and solves the equations of motion. 
Physically, the real scalar $X$ deforms the dilaton , axion and the warp factors as can be read  in eqs.(\ref{defi}). The six dimensional fields $F_2, F_3, F^i$ enter the expressions of the fluxes using eqs.(\ref{tendconfig}),  (\ref{eq4})
 and (\ref{eq:G5_def}) and fibre the two-sphere with the six dimensional space as expressed by eq.(\ref{eq4}).

\subsection{The explicit backgrounds}

Let us now explicitly write the uplift of the solution presented in section \ref{section3} to type IIB supergravity. For each of these backgrounds, the ten dimensional supergravity equations of motion and Bianchi identities have been checked at the AdS-fixed points.
\subsection{AdS$_4 \times H_2$ in Type IIB}\label{sec:ads4h2IIB}
We write here the background in Type IIB, describing the flow from AdS$_6$ to AdS$_4\times H_2$, as found in section \ref{AdS4H2}. We use the lift as described in  eqs.(\ref{tendconfig})- (\ref{eq:G5_def}).
\begin{eqnarray}
& & ds_{st}^2= f_1 \Big[    \frac{2\tilde{g}^2}{9}\left( e^{2 f(r)}(-dt^2+ dy_1^2+ dy_2^2 + e^{2 g(r)}dr^2) +  e^{2h(r)}\frac{(dz^2+ dx^2)}{z^2}   \right)  +\nonumber\\
& & \qquad \quad  f_2  d s^2 (\tilde{S}^2)+ f_3 (d\sigma^2+d\eta^2)\Big],\nonumber\\
& &B_2 = f_4\text{Vol}(\tilde{S}^2) + \frac{2}{9} \eta \frac{\cos\theta}{z^2} dx\wedge dz \, , \;\;\;C_2 = f_5\text{Vol}(\tilde{S}^2) +4  \partial_\sigma (\sigma V) \frac{\cos\theta}{z^2} dx\wedge dz  \, , \nonumber\\
& & G_5 = \frac{2}{3 X^2} \left( \frac{\sqrt{2} \gt}{3}\right)^4  e^{4 f(r)-2 h(r)+g(r)} d r \wedge d t \wedge d y_1 \wedge d y_2 \wedge d\left(\cos \theta  \sigma^2 \partial_\sigma V \right) \, ,\nonumber\\
& &  F_5 = 4 (G_5 +*_{10} G_5),\;\;\;\;C_0 = f_7,\;\;\;\;  e^{-2\Phi} = f_6.\label{AdS4H2inIIB}
\end{eqnarray}
We have used,
\begin{equation}
d s^2 (\tilde{S}^2)=d \theta^2 +\sin ^2\theta  (d\phi-A^3)^2 \, , \qquad \text{Vol}(\tilde{S}^2) = \sin \theta d \theta \wedge \left( d\phi - A^3 \right) \, ,\;\;\; A^3=\frac{dx}{z}.
\end{equation}
We also used the definitions of $f_1,....,f_7$ in eq.(\ref{defi}). These depend explicitly on the field $X(r)$. The expressions for $X(r)$, $f(r)$, $h(r)$ and $g(r)$ are read from section \ref{AdS4H2} and Appendix \ref{numericsAdS4}.

When considered at the fixed point of eq.(\ref{eq:AdS_4_fixed}), the configuration of eq.(\ref{AdS4H2inIIB}) solves all the equations of motion and describes an infinite family of AdS$_4$ backgrounds in Type IIB preserving four Poincar\'e supercharges. This family is labelled by a function $V(\sigma,\eta)$ solving eq.(\ref{diffeq}). These solutions should be part of the more generic backgrounds in \cite{Passias:2017yke}. At the fixed point, the backgrounds in eq \eqref{AdS4H2inIIB} are dual to an infinite family of three dimensional $\mathcal{N} = 2$ SCFTs.

Let us now move into a new family of backgrounds with an AdS$_3\times H_3$ factor in Type IIB.

\subsection{AdS$_3 \times H_3$ in Type IIB}\label{secads3IIB}

This case is a bit more complicated than the previous one. 
Below, we write the background in Type IIB, describing the flow from AdS$_6$ to AdS$_3\times H_3$, as found in section \ref{AdS3H3}. As above, we use the lift as described in eqs.(\ref{tendconfig})- (\ref{eq:G5_def}).
\begin{eqnarray}
& & ds_{st}^2= f_1 \Big[    \frac{2\tilde{g}^2}{9}\left(  e^{2 f(r)}(-dt^2 + dy^2+ e^{2g(r)} dr^2)+ e^{2 h(r)}\frac{(dx_1^2+ dx_2^2+dz^2)}{z^2} \right)  +\nonumber\\
& & \qquad \quad  f_2  d s^2 (\tilde{S}^2)+ f_3 (d\sigma^2+d\eta^2)\Big],\nonumber\\
& &B_2 = f_4\text{Vol}(\tilde{S}^2) - \frac{2}{9} \eta y^i F^i \, , \;\;\;C_2 = f_5\text{Vol}(\tilde{S}^2) -4  \partial_\sigma (\sigma V) y^i F^i \, , \nonumber\\
& & G_5 =  
 - \frac{2}{3 X^2} \left(\frac{\sqrt{2} \gt}{3}\right)^2 \frac{e^{3 f(r)-h(r)+g(r)}}{z} dr \wedge dt \wedge  d y_1 \wedge \bigg[ d x_2 \wedge d\left(\sin \theta \sin \phi \sigma^2  \partial_\sigma V \right)\nonumber \\
& & + d x_1 \wedge d\left(\cos \theta \sigma^2  \partial_\sigma V \right)- d z \wedge \left( d\left(\cos \phi \sin \theta \sigma^2  \partial_\sigma V \right) + \sigma^2  \partial_\sigma V (\cos \theta A_1 + \sin \theta \sin \phi A_2) \right) \bigg]\, ,\nonumber\\
& &  F_5 = 4 (G_5 +*_{10} G_5),\;\;\;\;C_0 = f_7,\;\;\;\;  e^{-2\Phi} = f_6.\label{AdS3H3inIIB}
\end{eqnarray}
We have used  the definitions in eqs.\eqref{configH3}-\eqref{eq:S2_embedding}  and 
\begin{eqnarray}
& & d s^2 (\tilde{S}^2)= (d \theta + \cos \phi A^1 )^2+\sin^2 \theta  (d \phi- \cot \theta \sin  \phi A^1- A^3)^2 \, , \\[2mm]
& & \text{Vol}(\tilde{S}^2) = \sin \theta (d \theta + \cos \phi A^1 ) \wedge (d \phi- \cot \theta \sin  \phi A^1- A^3) \, .\nonumber\\[2mm]
& & A^1= -\frac{dx_1}{z}, \;\; A^2=0,\;\; A^3= -\frac{dx_2}{z},\;\;\; F^1= \frac{dz\wedge dx_1}{z^2}, \;\; F^2= \frac{dx_2\wedge dx_1}{z^2},\;\; F^3= \frac{dz\wedge dx_2}{z^2},\nonumber
\end{eqnarray}
The functions $X(r)$, $f(r)$, $g(r)$, $h(r)$ solve the equations discussed in section \ref{AdS3H3} and Appendix \ref{numericsAdS3}.
For the particular values  in eq.(\ref{AdS3fixedpoint}) we checked  that the Type IIB equations of motion are satisfied. In this case the background in eq.(\ref{AdS3H3inIIB}) describes an infinite family of Type IIB SUSY backgrounds with an AdS$_3$ factor. There are dual to two dimensional $\mathcal{N} = (1,1)$ CFTs.

With the mechanism of the lift probably clear in the reader's mind, let us briefly discuss the new family of Type IIB backgrounds with an AdS$_2$ factor.

\subsection{AdS$_2 \times H_2 \times H_2$ in Type IIB}\label{secads2IIB}
The details needed to explicitly write  the ten-dimensional solution interpolating between AdS$_6$ and AdS$_2$ are the definitions of the fields $A^i$ in eq.\eqref{eq:AdS2xH2XH2_Ai}, the line and volume element for the two sphere as written below.
With these elements, we carefully follow eqs.(\ref{tendconfig})- (\ref{eq:G5_def}) and write the new infinite family of Type IIB solutions with an AdS$_2$ factor.
This family of backgrounds reads,
\begin{eqnarray}
& & ds_{st}^2= f_1 \Big[    \frac{2\tilde{g}^2}{9}\left( e^{2 f(r)}(-dt^2+ e^{2 g(r)}dr^2) +  e^{2h_1(r)}\frac{(dz_1^2+ dx_1^2)}{z_1^2} +  e^{2h_2(r)}\frac{(dz_2^2+ dx_2^2)}{z_2^2}  \right)  +\nonumber\\
& & \qquad \quad  f_2  d s^2 (\tilde{S}^2)+ f_3 (d\sigma^2+d\eta^2)\Big],\nonumber\\
& &B_2 = f_4\text{Vol}(\tilde{S}^2) - \frac{2\tilde{g}}{9} \sigma F_2 +   \frac{2}{9} \eta \cos\theta  F^3 \, , \;\;\;C_2 = f_5\text{Vol}(\tilde{S}^2) - 4 \tilde{g} (\sigma\partial_\eta V) F_2+       4  \partial_\sigma (\sigma V) \cos\theta  F^3 \, , \nonumber\\
& & G_5 =  - \left(\frac{\sqrt{2}}{3} \gt\right)^2 \left( \frac{1}{2\gt^2} F^3 \wedge F^3 \wedge *_2 d (\sigma ^2 \partial_\sigma V)+\frac{2 e^{2f}}{3 X^2} d r \wedge dt \wedge F^3 \wedge d(\cos \theta \sigma ^2 \partial_\sigma V) \right) ,\nonumber\\
& &  F_5 = 4 (G_5 +*_{10} G_5),\;\;\;\;C_0 = f_7,\;\;\;\;  e^{-2\Phi} = f_6.\label{AdS2H2H2inIIB}
\end{eqnarray}
We have used that,
\begin{eqnarray}
& & A^3= \frac{dx_1}{z_1} + \frac{dx_2}{z_2}, \qquad F^3= -\frac{dz_1\wedge dx_1}{z_1^2} - \frac{dz_2\wedge dx_2}{z_2^2},\nonumber\\
& & F_2= \frac{3}{2 \tilde{g}^3} X(r)^2 e^{2(f(r) - h_1(r) - h_2(r))} dt\wedge dr,\nonumber\\[2mm]
& & d s^2 (\tilde{S}^2)=d \theta^2 +\sin ^2\theta  (d\phi-A^3)^2 \, , \qquad \text{Vol}(\tilde{S}^2) = \sin \theta d \theta \wedge \left( d\phi - A^3 \right) \, . \nonumber 
\end{eqnarray}
In the case in which the fields $X(r)$, $e^{2f(r)}$, $e^{2h_1(r)}$ and $e^{2h_2(r)}$ take the values in eq.(\ref{AdS6toAdS2fixed}) we have checked that the configuration in eq.(\ref{AdS2H2H2inIIB}) solves the Type IIB equations of motion at the AdS$_2$ fixed point. In eq.(\ref{AdS2H2H2inIIB}), we have an infinite family of AdS$_2$ backgrounds in type IIB subject to the function $V(\sigma,\eta)$ solving the PDE (\ref{diffeq}).
\subsection{$R^{1,3}\times S^1$ in Type IIB}\label{sec:R13}
The lift to Type IIB of the six dimensional background in section \ref{BHWR} is specially simple. In fact, the function $X(r)$---the scalar in Romans supergravity is $X=1$. The fields $A_2, A_1$ and $A^i$ are all vanishing.
As a consequence of this, the line and volume element of $\tilde{S}^2$ are those of the rounded sphere, the Ramond five form vanish and we have,
\begin{eqnarray}
& & 
 ds_{st}^2= f_1 \Big[    \frac{2\tilde{g}^2}{9}\left(  e^{2\rho}\left(- d\tau^2+ d\vec{x}_3^2  \right) +\frac{d\rho^2}{g(\rho)}+ e^{2\rho} g(\rho) d\psi^2  \right)  +  f_2  d s^2 ({S}^2)+ f_3 (d\sigma^2+d\eta^2)\Big],\nonumber\\
& &B_2 = f_4\text{Vol}({S}^2) , \;\;\;C_2 = f_5\text{Vol}({S}^2)  \, ,\;\;\; C_0 = f_7,\;\;\;\;  e^{-2\Phi} = f_6.\label{R13S1inIIB}
\end{eqnarray}
The function $g(\rho)=  1-e^{5 (\rho_* -\rho)}$ with the constant $e^{2\rho_*}=\frac{4}{125}$ to avoid conical singularities. The functions $f_i(\sigma,\eta)$ are identical to the $\tilde{f}_i(\sigma,\eta)$ in (\ref{background1}), which in turns are equivalent to eq.(\ref{defi}) after setting $X=1$. All the functions $V(\sigma,\eta)$ solving eq.(\ref{diffeq}) define a solution in Type IIB with a four dimensional Minkowski space-time and a circle (breaking all SUSYs) that asymptotes to AdS$_6\times S^2\times \Sigma_2(\sigma,\eta)$. It is possible to construct another family of solutions using the non-SUSY AdS$_6$ vacuum with $3 X^4=1$. The stability of this can be tested using the methods of \cite{Dibitetto:2020csn}.

\section{Dual Field theories} \label{FTDsection}
In this section we discuss aspects of the field theories dual to the type IIB backgrounds in sections \ref{sec:ads4h2IIB}-\ref{sec:R13}. We briefly remind the reader some aspects of `twisted compactifications'. After that we discuss a meaningful observable, the holographic central charge. We define this quantity for the fixed-point solutions. Then, we extend this definition for a quantity  along the full flows from AdS$_6$.

\subsection{Field Theory duals to the Type IIB backgrounds}\label{sec:QFT}
The backgrounds in sections \ref{sec:ads4h2IIB}-\ref{sec:R13} have holographic duals. The logic to find these duals goes as follows. In a certain regime of the radial-coordinate $r$ (or  $\rho$  in section \ref{sec:R13}), the background asymptotes to AdS$_6$. The field $X(r)\sim 1$ in that regime, but we still have the fields $F^i$  (and $F_2$ in section \ref{secads2IIB}) switched on, breaking the $SO(2,5)$ isometry. From the ten dimensional perspective the AdS$_6$ isometries are broken both with the fluxes and the metric, which has a non trivial $S^2$ fibration. This is characteristic of twisted compactifications. In this case, an infinite family of 5d SCFTs (described by all the functions $V(\sigma,\eta)$ solving a Laplace equation) are compactified on curved manifolds: a hyperbolic plane $H_2$ in the case of section \ref{sec:ads4h2IIB}, a hyperbolic three space $H_3$ in section \ref{secads3IIB},  and a product $H_2\times H_2$ in section \ref{secads2IIB}. The twisted compactifications were introduced by Witten and nicely reviewed in \cite{Witten:1994ev}.  In the context of D-branes  the idea was developed by Berdshadsky, Sadov and Vafa in \cite{Bershadsky:1995qy}. Holographically, these compactifications were described in \cite{Maldacena:2000mw} and then worked out in detail in many examples, see \cite{Bobev:2017uzs},\cite{Acharya:2000mu} for a very small sample of such studies. For compactifications closely related to those discussed in sections \ref{AdS4H2}-\ref{AdS2H2H2}, see \cite{Kim:2019fsg}.

In the case at hand, we think our backgrounds as holographically describing the twisted compactification of five dimensional SCFTs. 
A case with good analytical control is the one of balanced linear quivers \cite{Legramandi:2021uds}. To be concrete, we stick to this case in what follows, but there results will apply for more generic SCFTs. These five dimensional linear quivers, reach a conformal fixed point at high energies. The SCFT is deformed by  operators describing the twisted compactification on a curved manifold ($H_2$, $H_3$ and $H_2\times H_2$ in the cases mentioned above). The non-trivial prediction of our geometries is that at low energies compared with the finite-size of the compact space, the field theories flow to an interacting  super conformal field theory in dimensions $(2+1)$, $(1+1)$ and $(0+1)$ respectively. Also, that some fraction of the original eight Poincar\'e supercharges is preserved at low energies. Other non-trivial predictions of the type IIB backgrounds are explored in section \ref{sec:holcent}.

The five dimensional SCFTs that we topologically-twist, describe the strongly coupled dynamics (at high energies) of a linear quiver field theory of the form

\begin{center}
	\begin{tikzpicture}
	\node (1) at (-4,0) [circle,draw,thick,minimum size=1.4cm] {N$_1$};
	\node (2) at (-2,0) [circle,draw,thick,minimum size=1.4cm] {N$_2$};
	\node (3) at (0,0)  {$\dots$};
	\node (5) at (4,0) [circle,draw,thick,minimum size=1.4cm] {N$_{P}$};
	\node (4) at (2,0) [circle,draw,thick,minimum size=1.4cm] {N$_{P-1}$};
	\draw[thick] (1) -- (2) -- (3) -- (4) -- (5);
	\node (1b) at (-4,-2) [rectangle,draw,thick,minimum size=1.2cm] {F$_1$};
	\node (2b) at (-2,-2) [rectangle,draw,thick,minimum size=1.2cm] {F$_2$};
	\node (3b) at (0,0)  {$\dots$};
	\node (5b) at (4,-2) [rectangle,draw,thick,minimum size=1.2cm] {F$_P$};
	\node (4b) at (2,-2) [rectangle,draw,thick,minimum size=1.2cm] {F$_{P-1}$};
	\draw[thick] (1) -- (1b);
	\draw[thick] (2) -- (2b);
	\draw[thick] (4) -- (4b);
	\draw[thick] (5) -- (5b);
	\end{tikzpicture}
\end{center}
The numbers $N_1, N_2,....,N_P$ and $F_1, ...., F_P$\footnote{In the balanced case they must satisfy the relation $F_i = 2 N_i - N_{i-1}-N_{i+1}$.} determine uniquely the function $V(\sigma,\eta)$. This is clearly explained in \cite{Legramandi:2021uds}, we refer the reader to that work. 

These UV conformal points  are deformed by relevant operators. The dimension of these operators can be read from the near-AdS$_6$ expansion of the gauged supergravity metric. These relevant operators (analogously the presence of the six dimensional gauge fields) topologically twist the 5d CFT and trigger a RG flow, that ends in a CFT$_3$, CFT$_2$, CFT$_1$ for the backgrounds in sections \ref{sec:ads4h2IIB}, \ref{secads3IIB}, \ref{secads2IIB} respectively.
An interesting quantity measuring the number of degrees of freedom (or the Free Energy) of these strongly coupled lower dimensional CFTs is defined in section \ref{choldefcal}. As we discuss below, the result is in terms of transcendental functions of the parameters defining the quiver, hence revealing the non-perturbative character of the CFT. An interesting result is that, in terms of the quiver parameters, the holographic central charge of the IR CFT is proportional to the UV one. This supports the picture advocated by Bobev--Crichigno \cite{Bobev:2017uzs}.

In contrast, the solution of section \ref{sec:R13} can be obtained by Wick rotating a Schwarzchild black hole in AdS$_6$. This background is also called the AdS-soliton. The solution in section \ref{sec:R13}, describes a
family of five dimensional SCFTs, that at the UV fixed point are compactified on a circle. After imposing periodic boundary conditions for the bosons and anti-periodic ones for the fermions (along a spatial circle), supersymmetry is broken, turning the scalars and fermions massive. In the perturbative spectrum, the gauge fields are massless. Our family of backgrounds describe the strong-dynamics of these systems.

Similar dynamics was exploited in \cite{Elander:2021wkc}-\cite{Elander:2020csd} with phenomenological purposes. It would be of interest to determine if the presence of bifundamental matter (even when massive) introduces any novel behaviour on observables.

\subsection{The holographic central charge}\label{sec:holcent}

%
%
%
%
%
%
%
%
%

In this section we calculate a quantity called holographic central charge. We start  in section \ref{choldefcal} by defining the quantity at conformal points and explicitly computing it in our AdS$_4$, AdS$_3$ and AdS$_2$ fixed point type IIB geometries.
This gives us information about the number of degrees of freedom in the strongly coupled CFT$_3$, CFT$_2$ and CFT$_1$ respectively.

After that, in section \ref{flowcentraldef} we present a quantity originally defined in  \cite{Bea:2015fja}. This quantity is inspired on the holographic central charge of \cite{Macpherson:2014eza}, but is applicable to geometries describing flows, like our AdS$_6 \to$ AdS$_{4,3,2}$.   We call this quantity the `flow central charge'. The main characteristic is that it is constant at both ends of the flow. As we explicitly show for our solutions, this quantity is monotonous along the flow and its value at the IR-fixed point is bigger than that at the UV-fixed point. 

Finally, in section \ref{flowAdS6R13} we compute the flow-central charge for the solution in section \ref{sec:R13} and interpret its result physically.
\subsubsection{The holographic central charge: definition and calculation at fixed points}\label{choldefcal}
First, we 
briefly remind the reader the formalism we use, developed in  \cite{Bea:2015fja},\cite{Macpherson:2014eza}. It can be briefly summarised as follows:
consider a $(d+1)$ dimensional QFT dual to a background, with metric  and dilaton of the form,
\begin{equation}
ds^2= a(r,\vec{\theta})\left[-dt^2 + d\vec{x}_d^2  + b(r) dr^2\right] + g_{ij}(r,\vec{\theta}) d\theta^id\theta^j ,\;\;\;\; e^{\Phi(r,\vec{\theta})}.\label{eq1x}
\end{equation}
We calculate a weighted version of the internal volume,
\begin{equation}
V_{\text{int}}(r)=\int d\vec{\theta} \sqrt{e^{-4 \Phi (r,\vec{\theta})} \det[g_{ij}] a^d(r,\vec{\theta})}.\label{eq2x}
\end{equation}
We define then a quantity $H(r)= V_{\text{int}}^2$ and compute the holographic central charge according to
\begin{equation}
c_\text{hol}= \frac{d^d}{G_N}b(r)^{\frac{d}{2}}\frac{H^{\frac{2d+1}{2}}}{(H')^d}.\label{cholx}
\end{equation}
Let us see this at work. First consider the fixed point solutions with AdS$_4$, then AdS$_3$ and finally AdS$_2$.
In these three fixed point solutions the full space time metric and dilaton are,
\begin{equation}
ds_{10,st}^2= f_1(\sigma,\eta)\left[ (\frac{2\tilde{g}^2}{9}) ds_6^2 + f_2(\sigma,\eta)d\tilde{\Omega}_2 + f_3(\sigma,\eta)(d\sigma^2+d\eta^2)\right],\;\;\; e^{-2\phi}= f_6(\sigma,\eta).\label{eq7x}
\end{equation}
For each of the fixed points we have
\begin{align}
& ds_{6,\text{AdS}_4}^2= e^{2f(r)}(- dt^2+dy_1^2+dy_2^2 + e^{2g(r)} dr^2)+ \frac{e^{2h(r)}}{z^2}(dz^2+ dx^2), \qquad \qquad \, \, \, d=2,\label{ads4-6}\\
&  ds_{6,\text{AdS}_3}^2= e^{2f(r)}(- dt^2+dy_1^2+ e^{2g(r)} dr^2)+ \frac{e^{2h(r)}}{z^2}(dz^2+ dx_1^2 + dx_2^2), \qquad \qquad \, \, \, d=1,\label{ads3-6}\\
&  ds_{6,\text{AdS}_2}^2\!=\! e^{2f(r)}(-\! dt^2 \! +\!e^{2g(r)} dr^2)+ \frac{e^{2h_1(r)}}{z_1^2}(dz_1^2\!+\! dx^2) +\frac{e^{2h_2(r)}}{z_2^2}(dz_2^2+ dx^2) , \qquad d=0.\label{ads2-6}
\end{align}
The functions
\begin{equation}
a(r,\vec{\theta})= \frac{2 \tilde{g}^2}{9}  f_1(\sigma,\eta) e^{2f(r)},\;\;\;\;\; b(r)= e^{2g(r)}.
\end{equation}
The `internal spaces' are
\begin{align}
& ds_{\text{int,\text{AdS}}_4}^2=f_1\left[ \frac{2 \tilde{g}^2}{9} \frac{e^{2h(r)}}{z^2}(dz^2+ dx^2) +  f_2d\tilde{\Omega}_2 + f_3(d\sigma^2+d\eta^2)   \right],\label{gintAdS4}\\
& ds_{\text{int,\text{AdS}}_3}^2=f_1\left[ \frac{2 \tilde{g}^2}{9} \frac{e^{2h(r)}}{z^2}(dz^2+ dx_1^2 +dx_2^2) +  f_2d\tilde{\Omega}_2 + f_3(d\sigma^2+d\eta^2)   \right],\label{gintAdS3}\\
& ds_{\text{int,\text{AdS}}_2}^2=f_1\left[   \frac{2 \tilde{g}^2}{9}  \frac{e^{2h_1(r)}}{z_1^2}(dz_1^2\!+\! dx_1^2) \!+\!\frac{e^{2h_2(r)}}{z_2^2}(dz_2^2\!+\! dx_2^2) \!+\! f_2d\tilde{\Omega}_2 + f_3(d\sigma^2+d\eta^2)   \right].
\end{align}
Calculating the combination in eq.(\ref{eq2x}) we have,
\begin{align}
& V_{\text{int,\text{AdS}}_4}= \left(\text{Vol} (S^2) \text{Vol} (H_2)\left(\frac{2\tilde{g}^2}{9} \right)^2\int d\sigma d\eta f_1^4  f_2  f_3  f_6 \right)\times e^{ 2f(r) + 2 h(r)  },\label{vintads4}\\
&  V_{\text{int,\text{AdS}}_3}= \left(\text{Vol} (S^2) \text{Vol} (H_3) \left(\frac{2\tilde{g}^2 }{ 9 } \right)^2\int d\sigma d\eta f_1^4 f_2  f_3  f_6 \right)\times e^{  f(r) + 3 h(r)  },\label{vintads3}\\
&  V_{\text{int,\text{AdS}}_2}= \left(\text{Vol} (S^2) \text{Vol} (H_{2,1}) \text{Vol}(H_{2,2}) \left(\frac{2\tilde{g}^2 }{ 9 } \right)^2\int d\sigma d\eta f_1^4 f_2  f_3  f_6 \right)\times e^{  2h_1(r) + 2 h_2(r)  }.\label{vintads2}
\end{align}
Using eq.(\ref{defi}), one finds
\begin{equation}
\int d\sigma d\eta f_1^4  f_2  f_3  f_6= \frac{2^6}{3}\int d\sigma d\eta \sigma^3 (\partial_\sigma V)(\partial^2_\eta V).\nonumber
\end{equation}
Now, if we define
\begin{eqnarray}
& & {\cal N}_{\text{AdS}_4}= \frac{2^6}{3} \text{Vol} (S^2) \text{Vol} (H_2)\left(\frac{2\tilde{g}^2}{9} \right)^2\int d\sigma d\eta \sigma^3 (\partial_\sigma V)(\partial^2_\eta V),\label{calNAdS4}\\
& & {\cal N}_{\text{AdS}_3}= \frac{2^6}{3} \text{Vol} (S^2) \text{Vol} (H_3) \left(\frac{2\tilde{g}^2}{9} \right)^2\int d\sigma d\eta \sigma^3 (\partial_\sigma V)(\partial^2_\eta V), \label{calNAdS3}\\
& & {\cal N}_{\text{AdS}_2}= \frac{2^6}{3} \text{Vol} (S^2) \text{Vol} (H_{2,1}) \text{Vol}(H_{2,2} ) \left(\frac{2\tilde{g}^2}{9} \right)^2\int d\sigma d\eta \sigma^3 (\partial_\sigma V)(\partial^2_\eta V),\label{calNAdS2}
\end{eqnarray}
we then have:
\begin{eqnarray}
& &   V_{\text{int,\text{AdS}}_4}= {\cal N}_{\text{AdS}_4} e^{2f+2h},\;\;\; V_{int,\text{AdS}_3}= {\cal N}_{\text{AdS}_3} e^{f+3h},\;\;\; V_{\text{int,\text{AdS}}_2}= {\cal N}_{\text{AdS}_2} e^{2h_1+2h_2},\nonumber\\[2mm]
& & H_{\text{AdS}_4}= {\cal N}_{\text{AdS}_4}^2 e^{4f+4h},\;\;\; H_{\text{AdS}_3}= {\cal N}_{\text{AdS}_3}^2 e^{2f+6h},\;\;\; H_{\text{AdS}_2}= {\cal N}_{\text{AdS}_2}^2 e^{4h_1+4h_2}.\nonumber
\end{eqnarray}  
Using eq.(\ref{cholx}) we find the holographic central charge for each fixed point to be,
\begin{eqnarray}
& & c_{\text{hol,AdS}_4}= \frac{{\cal N}_{\text{AdS}_4}}{4 G_N} \left( \frac{e^{f+g+h}}{f'+h'}\right)^2= \frac{{\cal N}_{\text{AdS}_4}}{4 G_N}\left(\frac{e^{f(r) + g(r)}}{f'(r)}\right)^2 e^{2h_0},\label{cholads4}\\
& & c_{\text{hol,AdS}_3}= \frac{{\cal N}_{\text{AdS}_3}}{ G_N} \left( \frac{e^{f+g+3h}}{f'+3h'}\right)= \frac{{\cal N}_{\text{AdS}_3}}{ G_N}\left( \frac{e^{f(r) + g(r)}}{f'(r)}   \right) e^{3h_0}.\label{cholads3}\\
& & c_{\text{hol,AdS}_2}= \frac{{\cal N}_{\text{AdS}_2}}{ G_N} \left( e^{2h_{1,0}+2h_{2,0}} \right),\label{cholads2}
\end{eqnarray}
We have denoted by $h_0$ the value of the warp function $h(r)$ at the fixed point, see eqs.(\ref{eq:AdS_4_fixed}), (\ref{AdS3fixedpoint}), (\ref{AdS6toAdS2fixed}) respectively. We used that, at the fixed point $h'(r)=0$, $f(r)=-\log \frac{r}{r_0}$, $g(r)=0$ (or we choose $g(r)=-f(r)=-\frac{r}{r_0}$). The value of $r_0$ can be computed using eqs.(\ref{eq:AdS_4_fixed}),(\ref{AdS3fixedpoint}).
For the particular case of AdS$_2$ with $d=0$ as indicated in eq.(\ref{ads2-6}) it is enough to compute with $V_{\text{int}}$ in eq.(\ref{vintads2}). This subtlety was discussed in \cite{Lozano:2020txg}.

Note that the factors ${\cal N}_{\text{AdS}_{4,3,2}}$ defined in eqs.(\ref{calNAdS4})-(\ref{calNAdS2}) involve, aside form the volumes of the compact internal space, an integral,
\begin{equation}
{\cal N}_{\text{AdS}_{6}}=\frac{2^6}{3} \text{Vol} (S^2)  \int d\sigma d\eta \sigma^3 (\partial_\sigma V)(\partial^2_\eta V).\label{integralI}
\end{equation}
This integral was found in  \cite{Legramandi:2021uds} when computing the holographic central charge/Free Energy of a general 5d SCFTs (our far UV SCFTs), see equation (3.7) in  \cite{Legramandi:2021uds}.
In the case of balanced linear quivers, this five dimensional holographic central charge can be computed explicitly and it has a transcendental dependence on the parameters defining the 5d quiver. See for example equation (3.21) in  \cite{Legramandi:2021uds}.

A way of physically understanding the expressions in eqs.(\ref{cholads4})-(\ref{cholads2}) is to think that the degrees of freedom of the five dimensional UV conformal theory
proportional to the quantity ${\cal N}_{\text{AdS}_{6}}$ in eq.(\ref{integralI}) gets `weighted' by the volume of the compactification manifold, represented by $e^{h_0}$, $\text{Vol} (H_i) $, etc. This confirms the picture advocated in 
\cite{Bobev:2017uzs}.

To close this section and motivate the next one, we observe by inspecting the generic expressions in eqs.(\ref{cholads4})-(\ref{cholads3}), that these do not present the fingerprint of  a UV fixed point. Namely, when $f(r)\sim h(r)\sim - \log r, g(r)=0$ we do not find a constant value for $c_\text{hol}$. In other words, the  AdS$_6$-UV fixed point of the flow is not captured by the quantity defined in eq.(\ref{cholx}). We lost this fixed point when we informed the theory that the space-dimension
of the dual CFT was either $d=2,1,0$ in eqs.(\ref{ads4-6})-(\ref{ads2-6}). The quantity defined in the next section remedies  this deficiency and shows the existence of both the IR and UV fixed points (except for the AdS$_2$ as we discuss). We call this quantity the `flow central charge'.
\subsubsection{The flow-central charge: definition and calculations}\label{flowcentraldef}
For the situation in which the QFT is actually a flow across dimensions (or the field theory lives in an anisotropic space time),
 we appeal to a slightly more elaborated formalism developed in \cite{Bea:2015fja}.
Adapting  the formalism of   \cite{Bea:2015fja} to our flows from AdS$_6$ backgrounds, leads us to consider metrics and dilaton of the form,
\begin{eqnarray}
& & ds^2= -\alpha_0 dt^2+ \alpha_1 dy_1^2 +\alpha_2 dy_2^2 + ....+ \alpha_d dy_d^2+
 \Pi_{i=1}^d (\alpha_1....\alpha_d)^{\frac{ 1 }{ d } } b(r) dr^2+\nonumber\\
 & & g_{ij} (d\theta^ i- A_1^i) (d\theta^j-A_1^j),\;\;\;\;\;\;\;\; e^{\Phi}.
\label{eq4x}
\end{eqnarray}
For the cases studied here, we set $d=4$ as we perform flows from AdS$_6$. We are interested in defining a quantity that is monotonous along the flow.
We define,
\begin{equation}
ds^2_{\text{int}}=  \alpha_1 dy_1^2 +\alpha_2 dy_2^2 + ....+ \alpha_d dy_d^2
 +g_{ij} (d\theta^ i- A_1^i) (d\theta^j-A_1^j), \qquad  e^{\Phi}.
\label{eq5x}
\end{equation}
We form the combination,
\begin{equation}
V_{\text{int}}= \int_{X} \sqrt{\det[g_{mn}]  e^{-4\Phi}}, \qquad H=V_{\text{int}}^2.\label{eq55x}
\end{equation}
The integral is over $X$  the manifold consisting of the internal space $g_{ij}$ and the dimensions `erased' by the RG-flow.
Then we define the holographic central charge along the flow as in eq.(\ref{cholx}). Namely, using that $d=4$ in all of our configurations,
\begin{equation}
c_{\text{flow}}= \frac{4^4}{G_N} b(r)^2 \left(\frac{H}{H'} \right)^4 H^{1/2}.\label{centralflowxx}
\end{equation}
For the flow metric interpolating between AdS$_6$ and AdS$_4 \times H_2$ in type IIB---see eqs.(\ref{AdS4H2inIIB})  and (\ref{ads4-6}), we have
\begin{eqnarray}
& & \alpha_1=\alpha_2= \frac{2\tilde{g}^2}{9} f_1(\sigma,\eta) e^{2f(r)},\;\;\; \alpha_3=\alpha_4= \frac{2\tilde{g}^2}{9} f_1(\sigma,\eta) e^{2h(r)},\nonumber\\
& & b(r)= e^{f(r) + 2g(r)- h(r) },\;\;\; V_{\text{int}}= {\cal N}_{\text{AdS}_4}  e^{2 f(r)+2h(r)},\;\;\; c_{\text{flow}}= \frac{{\cal N}_{\text{AdS}_4}}{ G_N}\left( \frac{e^{f+g}}{f'+h'}\right)^4.\label{cholflow4}
\end{eqnarray}
We have used the definition for ${\cal N}_{\text{AdS}_4}$ in eq.(\ref{calNAdS4}).
For the IR fixed point, when $f=-\log \frac{r}{r_0}$, $h=h_0$, $g=0$ and the UV fixed point with $f\sim h\sim -\log \frac{r}{r_0}$ and $g=0$ we find
\begin{equation}
c_\text{IR,flow}=\frac{{\cal N}_{\text{AdS}_4}}{ G_N} r_0^4, \;\;\;\;\; c_\text{UV,flow}=\frac{c_\text{IR,flow}}{16}.
\end{equation}
As before, the number $r_0$ can be calculated by imposing $f(r)=-\log \frac{r}{r_0}$ and $g(r)=0$ in eq.(\ref{eq:AdS_4_fixed}).
Notice that the value in the IR of the flow is proportional to that in eq.(\ref{cholads4}) and that the dependence on the `internal space' expressed by ${\cal N}_{\text{AdS}_4}$ is the same. This quantity does not indicate the number of degrees of freedom (as the IR of the flow contains more than the UV fixed point). But the quantity detects both fixed points. 

The case of the flow AdS$_6 \to$ AdS$_3\times H_3$ in type IIB represented by eqs.(\ref{AdS3H3inIIB}), (\ref{ads3-6}) works very much along the same lines, giving
\begin{eqnarray}
& &\alpha_1=  \frac{2\tilde{g}^2}{9} f_1(\sigma,\eta) e^{2f(r)},\;\;\;\alpha_2=\alpha_3=\alpha_4= \frac{2\tilde{g}^2}{9} f_1(\sigma,\eta) e^{2h(r)},\nonumber\\[2mm]
& &  b(r)= e^{\frac{3 f(r)}{2} + 2g(r)- \frac{3 h(r)}{2} }, \;\;\; V_{\text{int}}= {\cal N}_{\text{AdS}_3}  e^{ f(r)+3h(r)},\nonumber\\[2mm]
& & c_{\text{flow}}= \frac{16 {\cal N}_{\text{AdS}3}}{G_N} \left(\frac{e^{f+g} }{f'+3h'} \right)^4= \frac{16 {\cal N}_{\text{AdS}3}}{G_N} r_0^4 ,\;\;\;\;\; c_\text{IR,flow}= 256 ~ c_\text{UV,flow}.
\end{eqnarray}
As above, the number $r_0$ can be calculated by imposing $f(r)=-\log \frac{r}{r_0}$ and $g(r)=0$ in eq.(\ref{AdS3fixedpoint}).

\begin{figure}[!]
	\centering
	\includegraphics[scale=0.6]{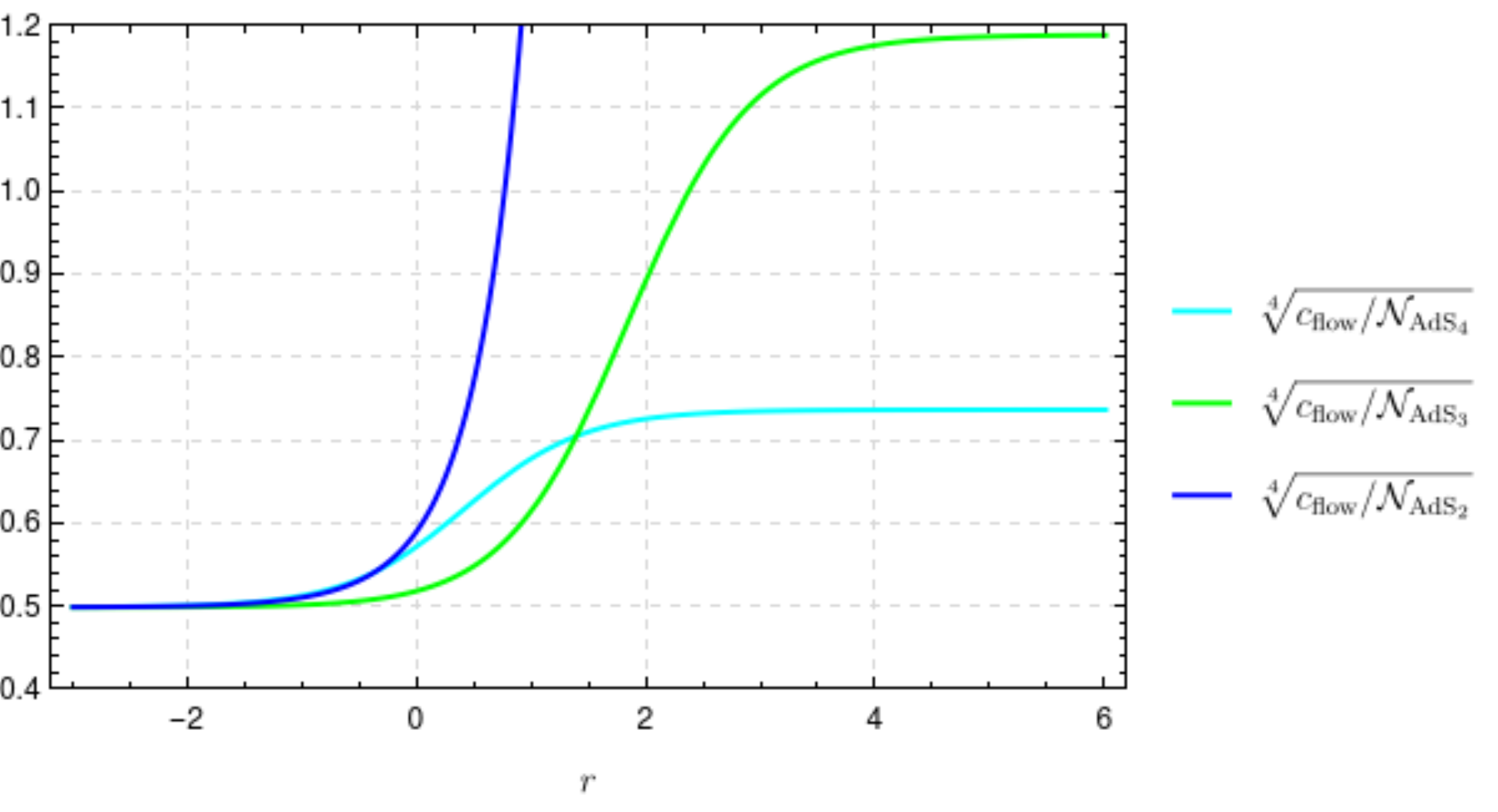}
	\caption{Plot of the flow central charge for the BPS solutions discussed in Figures \ref{fig:AdS4}, \ref{fig:AdS3} and \ref{fig:AdS2}. We have chosen $G_N=1$. The monotonicity of this quantity is clearly displayed.}
	\label{figuraX}
\end{figure}

More interesting is the case of the flow in eqs. (\ref{AdS2H2H2inIIB}), (\ref{ads2-6}). We obtain,
\begin{eqnarray}
& & b(r)= e^{2f(r) +2g(r) - h_1(r) -h_2(r)},\;\;\;\; V_{\text{int}}={\cal N}_{\text{AdS}_2} e^{2h_1(r) + 2 h_2(r)},\nonumber\\
& & c_{\text{flow}}= \frac{ {\cal N}_{\text{AdS}2}}{G_N} \left(\frac{e^{f+g } }{h_1'+h_2'} \right)^4.
\end{eqnarray}
This quantity detects the UV-AdS$_6$ fixed point, but  becomes divergent in the IR fixed point, when $h_1= h_2= h_0$ becomes constant. Hence we see that this flow-central charge should be handled with care. 
For the BPS solutions shown in Figures \ref{fig:AdS4},\ref{fig:AdS3},\ref{fig:AdS2}, we plot the flow central charge (setting $G_N=1$). The result is displayed in Figure \ref{figuraX}. 

In the next section, we study this same quantity for the flow of section \ref{sec:R13} and discuss the implications in Physics.

\subsubsection{The flow-central charge for the AdS$_6\to R^{1,3}\times S^1$ background}\label{flowAdS6R13}
We now consider the family of backgrounds discussed in section \ref{sec:R13}.
The metric and dilaton needed for this calculation are,
\begin{align}
&ds_{10,st}^2= f_1\left[e^{2\rho}\left(- d\tau^2+ d\vec{x}^2  \right) +\frac{d\rho^2}{g(\rho)}+ e^{2\rho} g(\rho) d\psi^2+ f_2d\Omega_2(\theta,\varphi) + f_3(d\sigma^2+d\eta^2) \right],\nonumber\\
&e^{-2\Phi}=972 e^{2\Phi_0}\frac{\sigma^2 \partial_\sigma V \partial^2_\eta V}{(3 \partial_\sigma V +\sigma \partial^2_\eta V)^2}\Lambda.\nonumber
\end{align}
The function $
g(\rho)= 1-e^{5 (\rho_* -\rho)}.
$
%
The functions $f_i(\sigma,\eta)$  and $\Lambda(\sigma,\eta)$ can be read from eq.(\ref{background1}), or from eq.(\ref{defi}), after setting $X=1$.

We can view this background as describing  a compactification on a circle or as a field theory that is five dimensional, but that has an anisotropy.
We use the formula developed in \cite{Bea:2015fja}, summarised in  eqs.(\ref{eq4x})-(\ref{centralflowxx}). 
To apply this requires the following assignation, 
 \begin{eqnarray}
 & & d=4,\;\;\;\;\; b(\rho)= e^{-2\rho} g(\rho)^{-\frac{5}{4}}\label{assignations}\\
 & & ds^2_\text{int}= f_1\Big[ e^{2\rho}\left( dx_1^2+dx_2^2+dx_3^2+ g(\rho)d\psi^2\right) + f_2 d\Omega_2 + f_3(d\sigma^2+d\eta^2)  \Big],\nonumber\\[2mm]
 & & V_{\text{int}}= \left(4\pi \int d\sigma d\eta e^{-2\Phi}f_1^4 f_2  f_3\right) e^{4\rho} \sqrt{g(\rho)}={\cal N}  e^{4\rho} \sqrt{g(\rho)}\nonumber\\[2mm]
 & & H= {\cal N}^2 e^{8\rho} g(\rho),\;\;\; H'= {\cal N}^2 e^{8\rho}(8 g(\rho) +g'(\rho)).\nonumber\\[2mm]
& & c_\text{hol}= \frac{ {\cal N} }{16 G_N} \frac{g^2}{(g+ \frac{g'}{8})^4 }=  \frac{ {\cal N} }{16 G_N} \frac{ \left(  1-  e^{-5 (\rho-\rho_*)} \right)^2}{\left(1-\frac{3}{8}  e^{-5 (\rho-\rho_*) } \right)^4}.\nonumber
 \end{eqnarray}
 This expression displays the expected behaviours. In the far IR, when $\rho\to\rho_*$, the value is zero indicating a gapped spectrum. At high energies, when $\rho\to\infty$, the result is that for a 5d SCFT. In fact, compare with eqs.(3.6)-(3.7) of \cite{Legramandi:2021uds}.
 There is an infinite number of linear quivers in 5d that upon compatification on a circle display a dynamics described by the background in section \ref{sec:R13}. These field theories differ in the choice of  ranks of colour and flavour groups
 for the linear quiver that in the far UV reaches a conformal point.
 
\begin{figure}[!]
	\centering
	\includegraphics[scale=0.6]{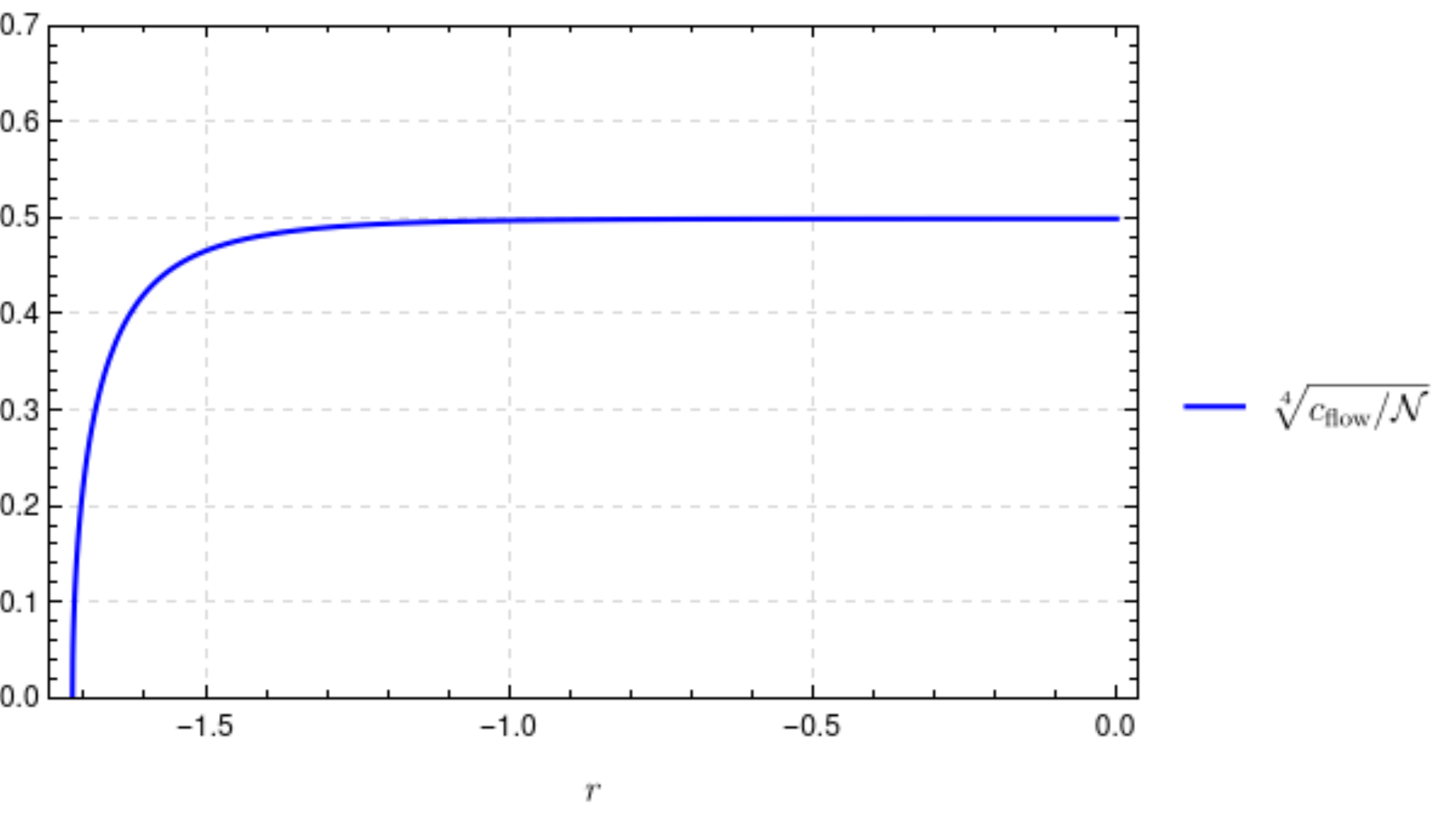}
	\caption{Plot of the flow central charge for the non-SUSY solution discussed in section \ref{sec:R13}. We have chosen $G_N=1$. The monotonicity of this quantity is clearly displayed.}
	\label{figuraXX}
\end{figure}

 Notice also the `separation' between the flow---represented by the $\rho$-dependent part of the central charge,  and the CFT structure ---represented by 
 $ \frac{ {\cal N} }{16 G_N}$ that this formula indicates. Similar `decoupling phenomenon' in different flows were observed in \cite{Faedo:2019cvr}. Following the formalism in \cite{Kol:2014nqa}, we can calculate the Entanglement Entropy in this case and find a behaviour qualitatively similar to the one  for the flow central charge.
The flow central charge for this QFT is plotted in Figure \ref{figuraXX}.

\section{Conclusions}\label{conclusions}
Let us start with a brief summary of the idea and achievements of this paper and close with  future lines of inquire.

After discussing four configurations in Romans' six dimensional $F_4$ gauged supergravity, we lift them to Type IIB generating new families of string backgrounds
with AdS$_4$, AdS$_3$ and AdS$_2$ that preserve SUSY. We also discussed  a family of solutions describing a circle compactification AdS$_6\to R^{1,3}\times S^1\times R_\rho$ that break SUSY.
All these backgrounds are parameterized by a function $V(\sigma,\eta)$, solving a Laplace equation. The compactifications to lower dimensional AdS$_{4,3,2}$ are dual to SCFTs in dimensions $(2+1)$, $(1+1)$ and $(0+1)$ dimensions. We identify the dual  field theory as the twisted compactification of a 5d `mother' SCFT on hyperbolic spaces of dimensions 2,3,4 respectively. The `mother' five dimensional SCFT preserves eight Poincar\'e SUSYs, it gets deformed by relevant operators that implement the twisted compactification. The kinematic data of the mother 5d SCFT is encoded in the function $V(\sigma, \eta)$ and the associated quiver and  matter content is inherited by the lower dimensional IR SCFT.

We study the number of degrees of freedom/Free Energy of the lower dimensional fixed points and emphasise  the non-perturbative nature of the result. We also define a monotonic quantity, characterising the flow. The non-SUSY compactification to a four dimensional quiver QFT with massive matter is also treated.

In the near future, it would be interesting to study different twisted compactifications and extend the picture displayed here to other Type II or M-theory backgrounds. Some steps in this direction are reported in \cite{Faedo:2019cvr}. The study of the Holographic Entanglement Entropy in these flow-geometries following  the formalism of \cite{Kol:2014nqa} might also give interesting information.

 The development of a map between SCFTs and supergravity backgrounds with less than half-BPS SUSY preserved is a worthwhile project. In this vein, it would be interesting (but not easy) to cast our infinite family of background with an AdS$_4$ factor in section \ref{sec:ads4h2IIB} in the language of \cite{Passias:2017yke}. There should be a way of relating our family of solutions with that of \cite{Bah:2018lyv}. The formalism developed in
\cite{Sacchi:2021afk} should apply for our backgrounds in section \ref{sec:ads4h2IIB}. There may be similar field theory elaborations for our backgrounds in sections \ref{secads3IIB} and \ref{secads2IIB}.
Obviously, a better understanding of the field theory dual to the background in section \ref{sec:R13} is very desirable, mostly to be used as a model with possible phenomenological interest.

We hope to report some progress in some of these lines above mentioned.

\section*{Acknowledgments:} Andrea Legramandi and Carlos Argentino Nunez wish to  thank the following colleagues for comments and discussions: Jeremias Aguilera-Damia, Diego Correa, Prem Kumar, Yolanda Lozano, Nicolo Petri, Guillermo Silva, Alessandro Tomasiello.
We are supported by STFC grant ST/T000813/1.

\appendix

\section{Conventions}\label{appendix1}


Our conventions agree with the one in \cite{Hong:2018amk} except for the definition of the Hodge dual of a $k$-form, which in our convention is given by
\begin{equation}
*(d x^{\mu_1} \wedge \dots \wedge d x ^{\mu_k}) = \frac{\sqrt{|g|}}{(D-k)} \epsilon^{\mu_1 \dots \mu_k}{}_{\nu_1 \dots \nu_{D-k}} d x^{\nu_1} \wedge \dots \wedge d x ^{\nu_{D-k}}
\end{equation}
where $D$ is the space-time dimension and 
\begin{equation}
\epsilon_{1 2 \dots D} = 1 \, .
\end{equation}
Notice that this definition differ respect to the one in \cite{Hong:2018amk} for the sign of the Hodge-dual of odd-forms; for example, we have
\begin{equation}
*_2 D y^i = - \epsilon_{ijk} y^j D y^k \, .
\end{equation}
This lead also to different signs in the kinetic terms in \eqref{eq:lagrangian6} and in the equations of motion \eqref{eqs6d}-\eqref{eqs6dlast}.

\section{Numerics}\label{sec:numerics}

In this section we give some details on the numerical analysis which leads to the flows from AdS$_6$ to lower dimensional AdS solutions. For the AdS$_d$ fixed points the idea will be the same: solve a linearized version of the BPS system close to the IR fixed point and use that analysis to give the initialization values for the numerical analysis. In the following, we will set $\gt = 2 / \sqrt{2}$ and we fix the parameterization invariance by setting $g = - f + \log X$.  We also define $h = - \frac12 \log Y$. 

\subsection{AdS$_4$ flow}\label{numericsAdS4}

After the redefinition in the previous above the BPS system \eqref{martinn1}-\eqref{martinn3} reads  
\begin{eqnarray}
& & X'+\frac{3}{4} X \left(X^{-2}-X^2-\frac{Y}{9}\right) = 0 \, , \label{eq:BPS_AdS41}\\
& & 3 X^2+X^{-2}-\frac{2 Y'}{Y}-Y = 0 \, ,\label{eq:BPS_AdS42} \\[2mm]
& & 12 f'+9 X^2+3 X^{-2}+Y = 0 \, . \label{eq:BPS_AdS43}
\end{eqnarray}
Using \eqref{eq:BPS_AdS41} as definition of $Y$, we can get a definition of $f$ in terms of $X$ and a second order ODE for $X$:
\begin{equation}
f' = -\frac1{X^2} - \frac{X'}{X}\, , \qquad 2 X''+\frac{10 X'^2}{X}+\left(\frac{14}{X^2}-24 X^2\right) X'+9 X^5-15 X+\frac{6}{X^3} = 0 \, .
\end{equation}
We can now solve this equation linearizing it close to the fixed point:
\begin{equation}
X = \sqrt[4]{\frac23} + \epsilon x(r) \, .
\end{equation}
At the first order in $\epsilon$, $x$ has to solve
\begin{equation}
2 x''-\sqrt{6} x'-12 x = 0
\end{equation}
which admits the solution
\begin{equation}
x = e^{-\frac{1}{2} \sqrt{\frac{3}{2}} \left(\sqrt{17}-1\right) r} \left(c_2 e^{\sqrt{\frac{51}{2}} r}+c_1\right) \, .
\end{equation}
Using this definition of $X$ we can derive the ones of $f$ and $Y$ at the first order in $\epsilon$:
\begin{eqnarray}
& & Y = 3 \sqrt{\frac{3}{2}}+ \epsilon \, 3 \sqrt[4]{54} e^{-\frac{1}{2} \sqrt{\frac{3}{2}} \left(\sqrt{17}-1\right) r} \left(\left(\sqrt{17}-4\right) c_2 e^{\sqrt{\frac{51}{2}} r}-\left(\sqrt{17}+4\right) c_1\right) \, , \\
& & f = -\sqrt{\frac{3}{2}} r+ \epsilon  \, \frac{1}{4} \sqrt[4]{\frac{3}{2}} e^{-\frac{1}{2} \sqrt{\frac{3}{2}} \left(\sqrt{17}-1\right) r}  \left(\left(\sqrt{17}-5\right) c_2 e^{\sqrt{\frac{51}{2}} r}-\left(\sqrt{17}+5\right) c_1\right) \, ,
\end{eqnarray}
where we have suppressed the integration constant of $f$ since it can be absorbed with a rescaling of the coordinates.

Now we can see that we have a superposition of two solutions, one which dominates for $r \gg 0$ and the other one for $r \ll 0$. If we set $c_1=0, r \ll 0$ or $c_2=0, r \gg 0$, we have that both the solution leads to the IR fixed point. In the first case however the numerics doesn't lead to a well behaving flow, so we consider the second possibility. Notice that if $r$ is big enough we don't need to consider $\epsilon$ very small, since the negative exponential is already a small correction to the IR contribution. For these reason we set $c_1 = \eps = 1$ and the initial condition for the plot in figure \ref{fig:AdS4} are given by the linearized expressions at $r=4$.

\subsection{AdS$_3$ flow}\label{numericsAdS3}

The discussion for the AdS$_3$ flow is very similar to the previous one.
After the redefinition the BPS system \eqref{BPSH3X}-\eqref{BPSH3h} reads  
\begin{eqnarray}
& & X'+\frac{3}{4} X \left(X^{-2}-X^2-\frac{Y}{3}\right) = 0 \, , \label{eq:BPS_AdS31}\\
& & 9 X^2+3 X^{-2}-\frac{6 Y'}{Y}-5Y = 0 \, ,\label{eq:BPS_AdS32} \\[2mm]
& & 4 f'+3 X^2+ X^{-2}+Y = 0\, . \label{eq:BPS_AdS33}
\end{eqnarray}
From \eqref{eq:BPS_AdS31} we can define $Y$ and the system reduces to:
\begin{equation}
f' = -\frac1{X^2} - \frac{X'}{X}\, , \qquad 6 X''+\frac{14 X'^2}{X}+\left(\frac{18}{X^2}-48 X^2\right) X'+18 X^5-27 X+\frac{9}{X^3} = 0 \, .
\end{equation}
We can now solve this equation linearizing it close to the fixed point:
\begin{equation}
X = \sqrt[4]{\frac12} + \epsilon x(r) \, .
\end{equation}
At the first order in $\epsilon$, $x$ has to solve
\begin{equation}
x''-\sqrt{2} x'-6 x = 0
\end{equation}
which admits the solution
\begin{equation}
x = e^{-\frac{\left(\sqrt{13}-1\right) r}{\sqrt{2}}} \left(c_2 e^{\sqrt{26} r}+c_1\right)\, .
\end{equation}
Using this definition of $X$ we can derive the ones of $f$ and $Y$ at the first order in $\epsilon$:
\begin{eqnarray}
& & Y = \frac{3}{\sqrt{2}}+ \epsilon \, 2^{3/4} e^{-\sqrt{7-\sqrt{13}} r} \left(\left(2 \sqrt{13}-7\right) c_2 e^{\sqrt{26} r}-\left(2 \sqrt{13}+7\right) c_1\right) \, , \\
& & f = -\sqrt{2} r+ \epsilon  \, \frac{1}{3} \sqrt[4]{2} e^{-\frac{\left(\sqrt{13}-1\right) r}{\sqrt{2}}} \left(\left(\sqrt{13}-4\right) c_2 e^{\sqrt{26} r}-\left(\sqrt{13}+4\right) c_1\right) \, .
\end{eqnarray}

Now we have again two solutions. The IR fixed point is at $c_2=0, r \gg 0$. The initial point for the numerical analysis in figure \ref{fig:AdS3} is given by the linear expressions at $c_1 = \epsilon = 1$ and $r=5$.

\subsection{AdS$_2$ flow } \label{numericsAdS2}

In this case we adopt the definition same definitions of $g$ and $\gt$ as before, moreover, as discussed in section \ref{AdS2H2H2}, we set $h_1 = h_2 = - \frac12 \log Y$. The BPS system therefore reduces to a set of three equations
\begin{eqnarray}
& & X'+\frac{3}{4} X \left(X^{-2}-X^2 \left(1- \frac{Y^2}{27}\right)-\frac{2Y}{9}\right) = 0 \, , \label{eq:BPS_AdS21}\\
& & \left(9 - \frac{Y^2}{3} \right)X^2+3 X^{-2}-\frac{6 Y'}{Y}-2Y = 0 \, ,\label{eq:BPS_AdS22} \\[2mm]
& & 12 f'+(9 + Y^2) X^2+3 X^{-2}+2Y = 0\, . \label{eq:BPS_AdS23}
\end{eqnarray}
We can use \eqref{eq:BPS_AdS21} to define $Y$, there are two possible solutions, but the one which leads to the AdS$_6$ fixed point is
\begin{equation}
Y = \frac{3}{X^2} (1-\sqrt{3 X^4 -4 X X'-2}) \, .
\end{equation}
Using this definition we have
\begin{align}
&f' = \frac{3 X X'+2 \sqrt{3 X^4-4 X X'-2}-3 X^4}{X^2} \, , \\
&X''-7 \frac{X'^2}{X}+3 (X^4-2 \sqrt{3 X^4-4 X X'-2}) \frac{X'}{X^2}+\frac{3 X^4-2}{X^3} (\sqrt{3 X^4-4 X X'-2}-1) = 0 \, .
\end{align}
Again, we linearize this equations near to the fixed point solution
\begin{equation}
X = \sqrt[4]{\frac12} + \epsilon x(r) 
\end{equation}
where
\begin{equation}
x = c_1 e^{-2 \sqrt{6} r}+c_2 e^{\sqrt{6} r} \, .
\end{equation}
From this definition it automatically follows that
\begin{equation}
Y = 3 \sqrt{\frac{3}{2}}-\sqrt[8]{3^{17} 2^7} \sqrt{c_1 \epsilon}  e^{-\sqrt{6} r} \, , \qquad f = - \sqrt{6} r -\sqrt[8]{3^{5} 2^{11}} \sqrt{c_1 \epsilon} e^{-\sqrt{6} r} \, .
\end{equation}
The IR fixed point is now obtained at $c_2=0, r \gg 0$. The initialization parameters for figure \ref{fig:AdS2} are given by the expressions above at $c_1 = \epsilon = 1$ and $r=5$.



\begin{thebibliography}{99}
 
\bibitem{Maldacena:1997re} 
  J.~M.~Maldacena,
  Int.\ J.\ Theor.\ Phys.\  {\bf 38}, 1113 (1999)
  [Adv.\ Theor.\ Math.\ Phys.\  {\bf 2}, 231 (1998)]
  [hep-th/9711200].

  

 



\bibitem{Lozano:2020txg}
Y.~Lozano, C.~Nunez, A.~Ramirez and S.~Speziali,
JHEP \textbf{03} (2021), 277
doi:10.1007/JHEP03(2021)277
[arXiv:2011.00005 [hep-th]].
%
\bibitem{Corbino:2020lzq}
D.~Corbino,
[arXiv:2004.12613 [hep-th]].


\bibitem{vanBeest:2020vlv}
M.~van Beest, S.~Cizel, S.~Schafer-Nameki and J.~Sparks,
SciPost Phys. \textbf{9}, no.3, 029 (2020)
[arXiv:2004.04020 [hep-th]].

\bibitem{Lozano:2020sae}
Y.~Lozano, C.~Nunez, A.~Ramirez and S.~Speziali,
JHEP \textbf{03} (2021), 145
doi:10.1007/JHEP03(2021)145
[arXiv:2011.13932 [hep-th]].
%
\bibitem{Lozano:2021rmk}
Y.~Lozano, C.~Nunez and A.~Ramirez,
JHEP \textbf{04} (2021), 110
doi:10.1007/JHEP04(2021)110
[arXiv:2101.04682 [hep-th]].

  
\bibitem{Couzens:2017way}
  C.~Couzens, C.~Lawrie, D.~Martelli, S.~Schafer-Nameki and J.~M.~Wong,
  [arXiv:1705.04679 [hep-th]].
  
  N.~T.~Macpherson,
  JHEP {\bf 1905} (2019) 089
  [arXiv:1812.10172 [hep-th]].
A.~Legramandi and N.~T.~Macpherson,
[arXiv:1912.10509 [hep-th]].
A.~Legramandi, N.~T.~Macpherson and G.~L.~Monaco,
[arXiv:2012.10507 [hep-th]]

 
  
\bibitem{Lozano:2019emq}
Y.~Lozano, N.~T.~Macpherson, C.~Nunez and A.~Ramirez,
JHEP \textbf{01}, 129 (2020)
[arXiv:1908.09851 [hep-th]].

 
\bibitem{Lozano:2019zvg}
Y.~Lozano, N.~T.~Macpherson, C.~Nunez and A.~Ramirez,
JHEP \textbf{01}, 140 (2020)
[arXiv:1909.10510 [hep-th]].


\bibitem{Lozano:2019jza}
Y.~Lozano, N.~T.~Macpherson, C.~Nunez and A.~Ramirez,
Phys. Rev. D \textbf{101}, no.2, 026014 (2020)
[arXiv:1909.09636 [hep-th]].


\bibitem{Lozano:2020bxo}
Y.~Lozano, C.~Nunez, A.~Ramirez and S.~Speziali,
JHEP \textbf{08} (2020), 118
[arXiv:2005.06561 [hep-th]].

\bibitem{Couzens:2021veb}
C.~Couzens, Y.~Lozano, N.~Petri and S.~Vandoren,
[arXiv:2109.10413 [hep-th]].

\bibitem{Lozano:2019ywa}
Y.~Lozano, N.~T.~Macpherson, C.~Nunez and A.~Ramirez,
JHEP \textbf{12}, 013 (2019)
[arXiv:1909.11669 [hep-th]].
  %


 

\bibitem{DHoker:2007hhe} 
  E.~D'Hoker, J.~Estes and M.~Gutperle,
  JHEP {\bf 0706}, 022 (2007)
  [arXiv:0705.0024 [hep-th]].
   E.~D'Hoker, J.~Estes, M.~Gutperle and D.~Krym,
  JHEP {\bf 0808}, 028 (2008)
  [arXiv:0806.0605 [hep-th]].


\bibitem{Assel:2011xz} 
  B.~Assel, C.~Bachas, J.~Estes and J.~Gomis,
  JHEP {\bf 1108}, 087 (2011)
  [arXiv:1106.4253 [hep-th]].
\bibitem{Bachas:2017wva}
C.~Bachas, M.~Bianchi and A.~Hanany,
JHEP \textbf{08}, 100 (2018)
[arXiv:1711.06722 [hep-th]].
C.~Bachas, I.~Lavdas and B.~Le Floch,
JHEP \textbf{10}, 253 (2019)
[arXiv:1905.06297 [hep-th]].
  
\bibitem{Assel:2012cj}
B.~Assel, C.~Bachas, J.~Estes and J.~Gomis,
JHEP \textbf{12} (2012), 044
doi:10.1007/JHEP12(2012)044
[arXiv:1210.2590 [hep-th]].
  
\bibitem{Lozano:2016wrs} 
  Y.~Lozano, N.~T.~Macpherson, J.~Montero and C.~Nunez,
  JHEP {\bf 1611}, 133 (2016)
  [arXiv:1609.09061 [hep-th]].

\bibitem{Coccia:2020wtk}
L.~Coccia and C.~F.~Uhlemann,
doi:10.1007/JHEP06(2021)038
[arXiv:2011.10050 [hep-th]].

\bibitem{Akhond:2021ffz}
M.~Akhond, A.~Legramandi and C.~Nunez,
[arXiv:2109.06193 [hep-th]].

  


\bibitem{Gaiotto:2009gz} 
  D.~Gaiotto and J.~Maldacena,
  JHEP {\bf 1210}, 189 (2012)
  [arXiv:0904.4466 [hep-th]].

\bibitem{ReidEdwards:2010qs} 
  R.~A.~Reid-Edwards and B.~Stefanski, jr.,
  Nucl.\ Phys.\ B {\bf 849}, 549 (2011)
  [arXiv:1011.0216 [hep-th]].
  
\bibitem{Aharony:2012tz} 
  O.~Aharony, L.~Berdichevsky and M.~Berkooz,
  JHEP {\bf 1208}, 131 (2012)
  [arXiv:1206.5916 [hep-th]].

\bibitem{Lozano:2016kum}
Y.~Lozano and C.~Nunez,
JHEP \textbf{05}, 107 (2016)
[arXiv:1603.04440 [hep-th]].

\bibitem{Nunez:2018qcj} 
  C.~Nunez, D.~Roychowdhury and D.~C.~Thompson,
  JHEP {\bf 1807}, 044 (2018)
  [arXiv:1804.08621 [hep-th]].
  C.~Nunez, D.~Roychowdhury, S.~Speziali and S.~Zacarias,
  Nucl.\ Phys.\ B {\bf 943}, 114617 (2019)
  [arXiv:1901.02888 [hep-th]].






\bibitem{DHoker:2016ujz} 
  E.~D'Hoker, M.~Gutperle, A.~Karch and C.~F.~Uhlemann,
  JHEP {\bf 1608}, 046 (2016)
  [arXiv:1606.01254 [hep-th]].
  

\bibitem{DHoker:2016ysh}
 E.~D'Hoker, M.~Gutperle and C.~F.~Uhlemann,
  Phys.\ Rev.\ Lett.\  {\bf 118}, no. 10, 101601 (2017)
  [arXiv:1611.09411 [hep-th]].
  E.~D'Hoker, M.~Gutperle and C.~F.~Uhlemann,
  JHEP {\bf 1705}, 131 (2017)
  [arXiv:1703.08186 [hep-th]].
  M.~Gutperle, C.~Marasinou, A.~Trivella and C.~F.~Uhlemann,
JHEP \textbf{09} (2017), 125
[arXiv:1705.01561 [hep-th]].
  E.~D'Hoker, M.~Gutperle and C.~F.~Uhlemann,
JHEP \textbf{11} (2017), 200
[arXiv:1706.00433 [hep-th]].
  
\bibitem{Gutperle:2018vdd} 
  M.~Gutperle, A.~Trivella and C.~F.~Uhlemann,
  JHEP {\bf 1804}, 135 (2018)
  [arXiv:1802.07274 [hep-th]].
M.~Fluder and C.~F.~Uhlemann,
  Phys.\ Rev.\ Lett.\  {\bf 121}, no. 17, 171603 (2018)
  [arXiv:1806.08374 [hep-th]].

\bibitem{Bergman:2018hin} 
  O.~Bergman, D.~Rodriguez-Gomez and C.~F.~Uhlemann,
  JHEP {\bf 1808}, 127 (2018)
  [arXiv:1806.07898 [hep-th]].
 
\bibitem{Lozano:2018pcp}
Y.~Lozano, N.~T.~Macpherson and J.~Montero,
JHEP \textbf{01} (2019), 116
[arXiv:1810.08093 [hep-th]].
   
   
\bibitem{Uhlemann:2019ypp}
  C.~F.~Uhlemann,
  arXiv:1909.01369 [hep-th].

 
 \bibitem{Uhlemann:2020bek}
C.~F.~Uhlemann,
JHEP \textbf{09} (2020), 145
doi:10.1007/JHEP09(2020)145
[arXiv:2006.01142 [hep-th]].


\bibitem{Legramandi:2021uds}
A.~Legramandi and C.~Nunez,
[arXiv:2104.11240 [hep-th]].



 
 
 






\bibitem{Apruzzi:2013yva}
  F.~Apruzzi, M.~Fazzi, D.~Rosa and A.~Tomasiello,
  JHEP {\bf 1404} (2014) 064
  [arXiv:1309.2949 [hep-th]].
  D.~Gaiotto and A.~Tomasiello,
  JHEP {\bf 1412} (2014) 003
  [arXiv:1404.0711 [hep-th]].

\bibitem{Cremonesi:2015bld}
  S.~Cremonesi and A.~Tomasiello,
  JHEP {\bf 1605} (2016) 031
  [arXiv:1512.02225 [hep-th]].

\bibitem{Nunez:2018ags}
  C.~Nunez, J.~M.~Penin, D.~Roychowdhury and J.~Van Gorsel,
  JHEP {\bf 1806} (2018) 078
  [arXiv:1802.04269 [hep-th]].
  K.~Filippas, C.~Nunez and J.~Van Gorsel,
  JHEP {\bf 1906}, 069 (2019)
  [arXiv:1901.08598 [hep-th]].

\bibitem{Bergman:2020bvi}
O.~Bergman, M.~Fazzi, D.~Rodriguez-Gomez and A.~Tomasiello,
[arXiv:2002.04036 [hep-th]].




\bibitem{Bah:2017wxp}
I.~Bah, A.~Passias and A.~Tomasiello,
JHEP \textbf{11} (2017), 050
doi:10.1007/JHEP11(2017)050
[arXiv:1704.07389 [hep-th]].
\bibitem{Passias:2017yke}
A.~Passias, G.~Solard and A.~Tomasiello,
JHEP \textbf{04} (2018), 005
[arXiv:1709.09669 [hep-th]].
\bibitem{Bobev:2019ore}
N.~Bobev, P.~Bomans and F.~F.~Gautason,
JHEP \textbf{06} (2020), 011
doi:10.1007/JHEP06(2020)011
[arXiv:1912.04779 [hep-th]].
\bibitem{Itsios:2017cew}
G.~Itsios, Y.~Lozano, J.~Montero and C.~Nunez,
JHEP \textbf{09} (2017), 038
doi:10.1007/JHEP09(2017)038
[arXiv:1705.09661 [hep-th]].
\bibitem{Bah:2018lyv}
I.~Bah, A.~Passias and P.~Weck,
JHEP \textbf{01} (2019), 058
[arXiv:1807.06031 [hep-th]].



%
\bibitem{Dibitetto:2017tve}
G.~Dibitetto and N.~Petri,
JHEP \textbf{12} (2017), 041
[arXiv:1707.06152 [hep-th]].
G.~Dibitetto and N.~Petri,
JHEP \textbf{01} (2018), 039
[arXiv:1707.06154 [hep-th]].

 
 
\bibitem{Romans:1985tw}
L.~J.~Romans,
Nucl. Phys. B \textbf{269} (1986), 691
doi:10.1016/0550-3213(86)90517-1
 
 \bibitem{Bobev:2017uzs}
 N.~Bobev and P.~M.~Crichigno,
 JHEP \textbf{12} (2017), 065
 doi:10.1007/JHEP12(2017)065
 [arXiv:1708.05052 [hep-th]].
 
%
\bibitem{Hong:2018amk}
J.~Hong, J.~T.~Liu and D.~R.~Mayerson,
JHEP \textbf{09} (2018), 140
doi:10.1007/JHEP09(2018)140
[arXiv:1808.04301 [hep-th]].




 
 	
\bibitem{Nunez:2001pt}
C.~Nunez, I.~Y.~Park, M.~Schvellinger and T.~A.~Tran,
JHEP \textbf{04} (2001), 025
doi:10.1088/1126-6708/2001/04/025
[arXiv:hep-th/0103080 [hep-th]].



\bibitem{Suh:2018tul}
M.~Suh,
JHEP \textbf{01} (2019), 035
doi:10.1007/JHEP01(2019)035
[arXiv:1809.03517 [hep-th]].
 
 



	\bibitem{Apruzzi:2018cvq}
	F.~Apruzzi, J.~C.~Geipel, A.~Legramandi, N.~T.~Macpherson and M.~Zagermann,
	Fortsch. Phys. \textbf{66}, no.3, 1800006 (2018)
	doi:10.1002/prop.201800006
	[arXiv:1801.00800 [hep-th]].
	
\bibitem{Cvetic:1999un}
M.~Cvetic, H.~Lu and C.~N.~Pope,
Phys. Rev. Lett. \textbf{83} (1999), 5226-5229
doi:10.1103/PhysRevLett.83.5226
[arXiv:hep-th/9906221 [hep-th]].
	


%
\bibitem{Dibitetto:2020csn}
G.~Dibitetto, N.~Petri and M.~Schillo,
JHEP \textbf{08} (2020), 040
doi:10.1007/JHEP08(2020)040
[arXiv:2002.01764 [hep-th]].

	
\bibitem{Witten:1994ev}
E.~Witten,
J. Math. Phys. \textbf{35} (1994), 5101-5135
doi:10.1063/1.530745
[arXiv:hep-th/9403195 [hep-th]].

\bibitem{Bershadsky:1995qy}
M.~Bershadsky, C.~Vafa and V.~Sadov,
Nucl. Phys. B \textbf{463} (1996), 420-434
doi:10.1016/0550-3213(96)00026-0
[arXiv:hep-th/9511222 [hep-th]].
	
	
\bibitem{Maldacena:2000mw}
J.~M.~Maldacena and C.~Nunez,
Int. J. Mod. Phys. A \textbf{16} (2001), 822-855
doi:10.1142/S0217751X01003937
[arXiv:hep-th/0007018 [hep-th]].

\bibitem{Acharya:2000mu}
B.~S.~Acharya, J.~P.~Gauntlett and N.~Kim,
Phys. Rev. D \textbf{63} (2001), 106003
doi:10.1103/PhysRevD.63.106003
[arXiv:hep-th/0011190 [hep-th]].
M.~Naka,
[arXiv:hep-th/0206141 [hep-th]].
J.~M.~Maldacena and C.~Nunez,
Phys. Rev. Lett. \textbf{86} (2001), 588-591
[arXiv:hep-th/0008001 [hep-th]].
H.~Nieder and Y.~Oz,
JHEP \textbf{03} (2001), 008
[arXiv:hep-th/0011288 [hep-th]].
J.~P.~Gauntlett, N.~Kim and D.~Waldram,
Phys. Rev. D \textbf{63} (2001), 126001
[arXiv:hep-th/0012195 [hep-th]].
J.~D.~Edelstein and C.~Nunez,
JHEP \textbf{04} (2001), 028
[arXiv:hep-th/0103167 [hep-th]].
J.~Gomis,
Nucl. Phys. B \textbf{606} (2001), 3-17
[arXiv:hep-th/0103115 [hep-th]].
I.~Bah, C.~Beem, N.~Bobev and B.~Wecht,
JHEP \textbf{06} (2012), 005
doi:10.1007/JHEP06(2012)005
[arXiv:1203.0303 [hep-th]].
	
	%
\bibitem{Kim:2019fsg}
N.~Kim and M.~Shim,
Nucl. Phys. B \textbf{951} (2020), 114882
[arXiv:1909.01534 [hep-th]].
P.~Karndumri,
JHEP \textbf{01} (2013), 134
[arXiv:1210.8064 [hep-th]].
	
\bibitem{Elander:2021wkc}
D.~Elander, M.~Piai and J.~Roughley,
Phys. Rev. D \textbf{104} (2021) no.4, 046003
doi:10.1103/PhysRevD.104.046003
[arXiv:2103.06721 [hep-th]].
D.~Elander, M.~Piai and J.~Roughley,
Phys. Rev. D \textbf{103} (2021) no.4, 046009
doi:10.1103/PhysRevD.103.046009
[arXiv:2011.07049 [hep-th]].
%
D.~Elander, M.~Piai and J.~Roughley,
Phys. Rev. D \textbf{103} (2021), 106018
doi:10.1103/PhysRevD.103.106018
[arXiv:2010.04100 [hep-th]].

\bibitem{Elander:2020csd}
D.~Elander, M.~Piai and J.~Roughley,
JHEP \textbf{06} (2020), 177
[erratum: JHEP \textbf{12} (2020), 109]
doi:10.1007/JHEP06(2020)177
[arXiv:2004.05656 [hep-th]].
	
\bibitem{Bea:2015fja}
Y.~Bea, J.~D.~Edelstein, G.~Itsios, K.~S.~Kooner, C.~Nunez, D.~Schofield and J.~A.~Sierra-Garcia,
JHEP \textbf{05}, 062 (2015)
doi:10.1007/JHEP05(2015)062
[arXiv:1503.07527 [hep-th]].
	
	
	
	\bibitem{Macpherson:2014eza}
	N.~T.~Macpherson, C.~Nunez, L.~A.~Pando Zayas, V.~G.~J.~Rodgers and C.~A.~Whiting,
	JHEP \textbf{02}, 040 (2015)
	doi:10.1007/JHEP02(2015)040
	[arXiv:1410.2650 [hep-th]].
	

	
	
	

\bibitem{Faedo:2019cvr}
A.~F.~Faedo, C.~Nunez and C.~Rosen,
JHEP \textbf{03} (2020), 080
doi:10.1007/JHEP03(2020)080
[arXiv:1912.13516 [hep-th]].
	  
\bibitem{Kol:2014nqa}
U.~Kol, C.~Nunez, D.~Schofield, J.~Sonnenschein and M.~Warschawski,
JHEP \textbf{06} (2014), 005
doi:10.1007/JHEP06(2014)005
[arXiv:1403.2721 [hep-th]].
	
	

%
\bibitem{Sacchi:2021afk}
M.~Sacchi, O.~Sela and G.~Zafrir,
[arXiv:2105.01497 [hep-th]].
	
\end{thebibliography}
\end{document}